# A cost-effective strategy of enhancing machine learning potentials by transfer learning from a multicomponent dataset on ænet-PyTorch

*An Niza El Aisnada[a,b], Kajjana Boonpalit[b,d], Robin van der Kruit[b], Koen M. Draijer[b], Jon Lopez-Zorrilla[c], Masahiro Miyauchi[a], Akira Yamaguchi[a,e], Nongnuch Artrith[b*]*

[a] Department of Materials Science and Engineering, School of Materials and Chemical Technology, Tokyo Institute of Technology, 2−12−1 Ookayama, Meguro-ku, Tokyo 152−8552, Japan

[b] Materials Chemistry and Catalysis, Debye Institute for Nanomaterials Science, Utrecht University, 3584 CG Utrecht, The Netherlands

[c] Physics Department, University of the Basque Country (UPV/EHU), Leioa, Basque Country, Leioa, Spain

[d] School of Information Science and Technology, Vidyasirimedhi Institute of Science and Technology, Rayong, Thailand

[e] Biofunctional Catalyst Research Team, RIKEN Center for Sustainable Resource Science, 2-1 Hirosawa, Wako, Saitama 351-0198, Japan

**Abstract**

Machine learning potentials (MLPs) offer efficient and accurate material simulations, but constructing the reference ab initio database remains a significant challenge, particularly for catalyst-adsorbate systems. Training an MLP with a small dataset can lead to overfitting, thus limiting its practical applications. This study explores the feasibility of developing computationally cost-effective and accurate MLPs for catalyst-adsorbate systems with a limited number of ab initio references by leveraging a transfer learning strategy from subsets of a comprehensive public database. Using the Open Catalyst Project 2020 (OC20)—a dataset closely related to our system of interest—we pre-trained MLP models on OC20 subsets using the ænet-PyTorch framework. We compared several strategies for database subset selection. Our findings indicate that MLPs constructed via transfer learning exhibit better generalizability than those constructed from scratch, as demonstrated by the consistency in the dynamics simulations. Remarkably, transfer learning enhances the stability and accuracy of MLPs for the CuAu/$H_2O$ system with approximately 600 reference data points. This approach achieved excellent extrapolation performance in molecular dynamics (MD) simulations for the larger CuAu/$6H_2O$ system, sustaining up to 250 ps, whereas MLPs without transfer learning lasted less than 50 ps. We also examine the potential limitations of this strategy. This work proposes an alternative, cost-effective approach for constructing MLPs for the challenging simulation of catalytic systems. Finally, we anticipate that this methodology will pave the way for broader applications in material science and catalysis research, facilitating more efficient and accurate simulations across various systems.

1. **Introduction**

Machine learning (ML) has rapidly advanced due to improvements in computational power, the availability of vast datasets, and algorithmic innovations. ML methods have become pivotal for material simulations in chemistry, physics, and material science. Traditional ab initio methods, including Monte Carlo and Molecular Dynamics, face limitations in simulating large-scale material systems, often missing long-range interaction phenomena. Consequently, ML methods have been developed to emulate the precision of interatomic potentials derived from these accurate ab initio methods, leading to efficient ML-based interatomic potentials (MLPs).

Artificial Neural Networks (ANN) algorithms within MLP model development have gained popularity due to their capacity to discern complex relationships within various material systems by learning nonlinear patterns directly from data[1,2]. Current advancements have further challenged the field by integrating force components into MLPs. The inclusion of force components is crucial because it enhances the accuracy and generalizability of the models, enabling better prediction of physical properties such as elastic and vibrational characteristics. This integration leads to improved stability in simulations and more efficient training processes[3]. Despite significant progress in developing MLPs, constructing the ab initio database as the training reference remains a significant challenge, especially for complex systems like catalyst-adsorbate interactions.

One major challenge in training MLPs is dataset scarcity. Constructing a diverse custom dataset for a specific chemical system requires substantial effort and resources, and training an MLP with a small dataset can lead to overfitting, thus limiting its practical applications. To address this limitation, pre-trained models can be fine-tuned on the target dataset, resulting in more generalized MLPs than training from scratch. This approach facilitates the development of more specialized applications of MLPs.

Recent advancements have established various open-source multicomponent material databases intended for constructing machine learning potentials for various purposes[3–9]. However, the sheer volume of data in these databases necessitates substantial computational resources, presenting a significant challenge. Therefore, we aim to investigate the feasibility of utilizing MLPs from a subset of a large multi-component dataset and applying transfer learning to construct MLPs for our material of interest.

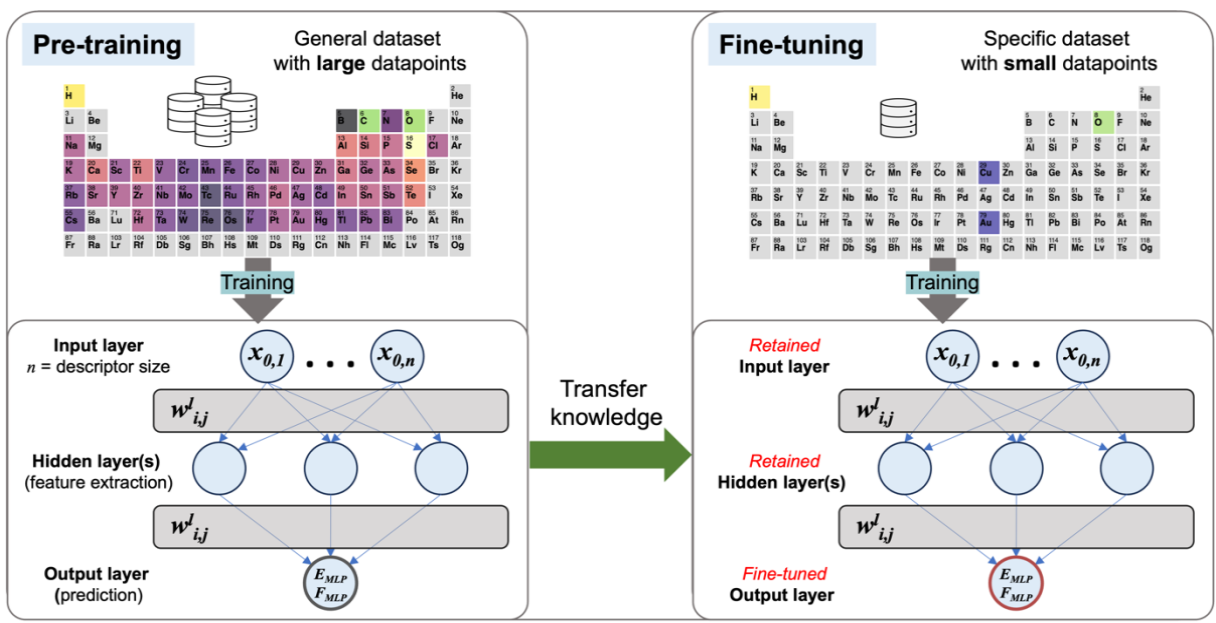

**Figure 1**. The transfer learning concept uses an ANN-based MLP model. This process is principally a knowledge transfer from pre-training to fine-tuning models. The pre-training model is trained on a large dataset with vast learning information, including the atomic environment. The transfer knowledge process involves retaining the pre-training information, including the input and hidden layers. The learned information is then adapted (fine-tuned) to train a specific database to improve predictions. The process efficiently transfers knowledge from a general to a specific context, enhancing model accuracy with limited data.

Transfer learning has emerged as a powerful technique to overcome the limitations of database construction and enhance the transferability of MLPs[10]. It involves training models on large, diverse datasets and then fine-tuning them on smaller, specifically targeted datasets (Figure 1). This process leverages the broad knowledge from large datasets to improve predictions on smaller ones. This approach is particularly valuable in material simulations, where obtaining large, high-quality datasets for every specific system presents a substantial challenge. Compared to routine MLP construction, which requires extensive data collection and training from scratch for each specific system, transfer learning significantly reduces the data and computational resources needed. By leveraging large, existing datasets, transfer learning accelerates development and enhances model generalization across different material systems. This efficient and cost-effective strategy is precious in fields like electrochemistry, catalysis, and materials design, where rapid and accurate simulations are crucial.

In this report, we explore cost-efficient transfer learning using subsets from the Open Catalyst 2020 (OC20) database[11]. The database contains a comprehensive chemical environment of 2 million catalyst-adsorbate systems that are similar, though not exactly the same, as our target applications. The OC20 dataset primarily includes periodic structures, while we are interested in cluster systems, which are finite and lack long-range periodicity. We employ two strategies for subset selection: (1) randomly selecting data that closely represent the original database, and (2) filtering by the chemical environment close to our applications. We pre-trained the MLP models using these subsets, then transferred and fine-tuned the models for our specific applications, which have smaller ab initio reference datasets. We evaluate the performance of transfer-learning MLPs using molecular dynamics (MD) simulations, in comparison with MLPs constructed directly from our dataset (without transfer learning). All MLP work is carried out by the Atomic Energy Network (ænet) software package. Specifically, we constructed the MLP using the ænet-PyTorch extension and performed the MD simulation using the ænet-LAMMPS extension. This work seeks a cost-effective and efficient way to construct MLPs for complex catalyst-adsorbate systems.

## 1.1 Machine learning interatomic potential (MLP) using ANN model

The success of Artificial Neural Network (ANN)-based Machine Learning Potentials (MLPs) largely hinges on their ability to model complex, nonlinear relationships within atomic systems[6–8,12]. These models effectively learn intricate patterns of atomic interactions regardless of the number of atoms, making them highly accurate for predicting material properties. ANNs operate by taking input features and processing them through multiple layers of interconnected 'neurons,' each applying a mathematical transformation. The output layer then delivers the final prediction.

In constructing MLPs, the ANN model interpolates atomic energy from reference *ab initio* calculations, such as Density Functional Theory (DFT). Behler and Parrinello introduced a method to harness the versatility of ANNs in creating practical and reusable interatomic potentials[2]. This approach, later expanded by Artrith, Morawietz, and Behler for multiple chemical species[13], involves decomposing the total energy $E(\sigma)$ of an atomic structure (σ) into individual atomic energy contributions $E_i$:

$$E^{ANN}(\{\sigma^{\{i\}}\}) = \sum_{i}^{N_{atom}} E_i(\sigma^{(i)}) \qquad (1)$$

The key to the success of ANN-based MLPs lies in their ability to predict energies and forces for new structures regardless of atom count, although they are confined to the chemical species present in the training data. This is accomplished by summing up the energy contributions from each atom, where each contribution $E_i$ is determined by the network corresponding to its element. The assumption here is that the energy contribution of each atom $i$ depends solely on its local environment, denoted as $\sigma_i$.

By defining these environments through atomic fingerprints or descriptors and training on the total energy of the system, the resulting potential gains generalizability, independent of the atom count. These descriptors must meet certain symmetry conditions with respect to atom exchanges, rotations, and translations of structures, as well as the smoothness of descriptor functions. Descriptors play a critical role in ANNs by representing the atomic environment in a format that the neural network can process. They transform the atomic positions and types into numerical values that encapsulate the essential features of the system[14,15].

## 1.2 ænet-Pytorch

ænet-PyTorch is a powerful implementation designed to train artificial neural network (ANN)-based machine learning potentials (MLPs) efficiently using the PyTorch framework. Developed as an extension of the atomic energy network (ænet), ænet-PyTorch leverages the computational power of graphics processing units (GPUs) to significantly accelerate the training process, making it feasible to include both energy and force data in the training, which enhances the accuracy and stability of the resulting MLPs.

The original ænet code, written in Fortran, supports the use of various descriptors to represent the local atomic environments. Among these, the Chebyshev descriptors introduced by Artrith et al. have shown great promise. These descriptors utilize Chebyshev polynomials to create a set of numerical values that describe the atomic environment around each atom, capturing both radial and angular dependencies. The Chebyshev descriptors are particularly efficient in handling systems with multiple chemical species because the size of the descriptor does not scale with the number of species.

The Chebyshev descriptor $T_n$ for an atom i can be represented as:

$$T_n^{cheb}(r_{ij}) = \sum_{n=0}^{N-1} a_n \cos\left(\frac{n\pi r_{ij}}{R_c}\right) \tag{2}$$

where $r_{ij}$ is the distance between atoms $i$ and $j$, $R_c$ is the cutoff radius, and $a_n$ are the coefficients of the Chebyshev polynomials. This formulation ensures that the descriptors are continuous and differentiable, which is crucial for the stability and accuracy of the MLPs.

ænet-PyTorch integrates these descriptors into the training process, allowing the neural network to learn from both energy and force data. The training workflow involves grouping atoms of the same species and processing their descriptors in parallel, significantly speeding up the computation. By utilizing PyTorch's built-in routines for tensor operations, the implementation achieves great scalability and performance, especially on GPUs. Using force information in the training process is essential for improving the generalizability and transferability of the MLPs. In ænet-PyTorch, the loss function combines both energy and force errors, weighted by a parameter α, allowing the model to balance the accuracy between energy and force predictions. The objective function $\mathcal{L}_{EF}$ is defined as:

$$\mathcal{L}_{EF} = (1 - \alpha)\mathcal{L}_E + \mathcal{L}_F \qquad (3)$$

where $\mathcal{L}_E$ and $\mathcal{L}_F$ are the root mean squared errors (RMSE) for energy and forces, respectively. This approach ensures that the trained MLPs can accurately predict atomic forces, which are critical for molecular dynamics simulations and other applications that require detailed atomic-level insights. By incorporating advanced descriptors like the Chebyshev polynomials and leveraging the computational power of GPUs, ænet-PyTorch facilitates the creation of highly accurate and transferable MLPs. This makes it an invaluable tool for researchers working on complex material simulations, including those involving catalyst-adsorbate interactions, as it enables the efficient and cost-effective training of machine learning potentials.

## 2. Methods

### 2.1 Selecting subsets from the OC20 Dataset

The Open Catalyst 2020 (OC20) database is a comprehensive resource for studying catalyst-adsorbate systems, offering relaxed atomic structures and energy and force data from DFT calculations. It includes diverse catalyst structures with adsorbates containing carbon, hydrogen, nitrogen, and oxygen. For cost-effective transfer learning, we compared two subset selection strategies for pre-training models: (1) randomly selecting representative data and (2) filtering by the chemical environment relevant to our applications. Specifically, for our transfer learning application involving copper-gold alloy clusters and water (CuAu/H₂O) systems, the second strategy focused on selecting structures containing "Cu" and "Au" to construct and utilize MLPs efficiently.

Table 1. Subset of OC20 for the pre-trained MLP models.

| OC20 Subset | Data points | Selection process |
|---|---|---|
| Random | 200K | Based on the closest representation of dataset in terms of energy, forces, number of atom/structure, and element distribution. |
| CuAu-54 | 170K | Filter any structures that contain at least "Cu", "Au", and "Cu + Au". |
| CuAu-11 | 4K | Filter any structures that contain at least "Cu + Au". |

## 2.2 MLP construction in ænet-PyTorch

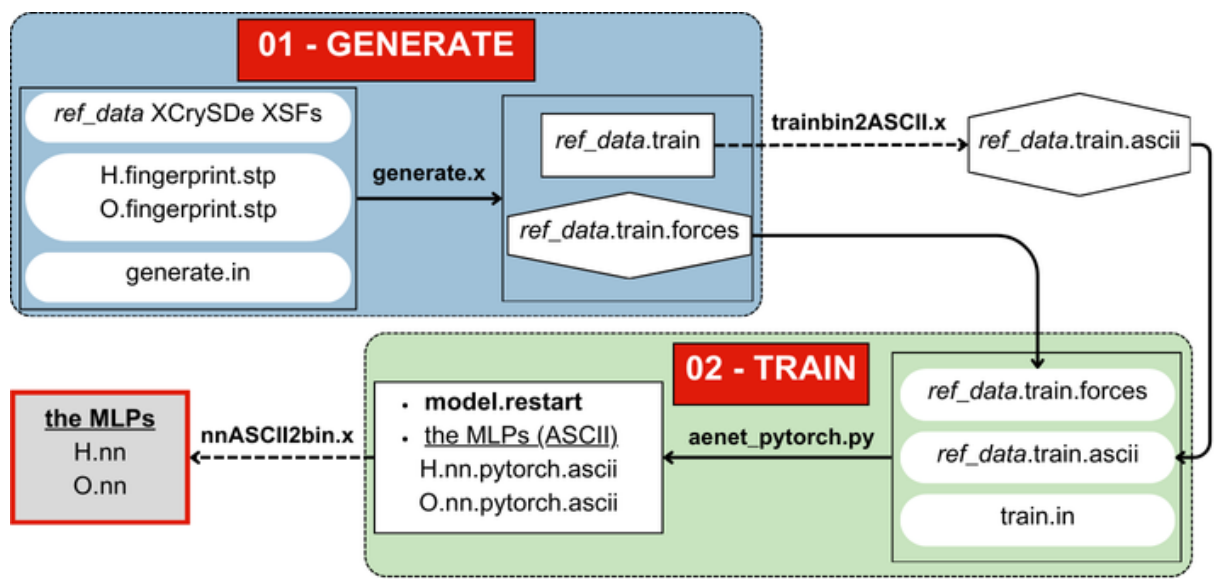

**Figure 2.** Technical workflow of constructing machine learning potential (MLP) using ænet-PyTorch for compounds containing Hydrogen (H) and Oxygen (O) as an example.

In general, constructing MLP in ænet consists of two main steps: (i) generation of the training set, which processes a compilation of the references dataset into a training set file; (ii) MLP model training, a process to construct the MLP models based on the generated training set file. In the ænet-PyTorch extension[16], the training of forces is possible. Thus, the generation of the training set will result in 2 files (energy set and forces set) if force training is enabled. Overall, the MLP construction in ænet-PyTorch is summarized in Figure 2.

### 2.2.1 Generation of the training set using generate.x and trainbin2ASCII.x tools

In generating the training set, one should provide input files, including the structural energy and force reference data in XCrySDen Structure Format (XSF), invariant basis or structural fingerprint for each element in the training set, *element.fingerprint.stp* files, and principle input file of *generate.in*. After executing *generate.x*, this process will result in the training set in the binary format *ref_data.train*. When the force is included in the training set, this process will also result in an additional file *ref_data.train.forces*. Further, we convert *ref_data.train* into a Python-readable ASCII format (*ref_data.train.ascii*) training set by executing *trainbin2ASCII.x* tool.

### 2.2.2 Construction of MLP using ænet_pytorch.py tool

In the routine of MLP construction using ænet, the *train.x* tool is used. However, in the ænet-PyTorch extension, we use the *aenet_pytorch.py* tool and execute it using Python. In this process, the input files for this step include the MLP training parameter in *train.in* file along with *ref_data.train.ascii* and *ref_data.train.forces* from the previous step. We implement a dataset split into 90% training and 10% test data during the training process. The training process results in the standard outputs, including the MLP for each element in the training set in ASCII format and *model.restart* as the training check-point. Finally, the executable binary MLPs (*element.nn*) for ænet-LAMMPS can be obtained by executing *nnASCII2bin.x* tool.

*2.2.3 Hyperparameter tuning*

To achieve optimal accuracy for the MLP models, we experimented with various hyperparameters during the generation of the training set and MLP training. For the generation of the training set, we focused on optimizing the size of descriptors and the cutoff radius dimensions. This involved adjusting hyperparameters within the fingerprint setup for each element (*.stp files). We explored descriptor sizes ranging from 36 to 68 by varying the degrees of radial and angular expansion functions ('*radial_N*' and '*angular_N*'). We surveyed the optimal radial cutoff radius ('*radial_Rc*') to be between 6 and 14 Å, and the angular cutoff radius ('*angular_Rc*') to be between 4 and 8.

Furthermore, to select the optimal hyperparameters for training, we evaluated various options, including optimization methods (Adam, Adadelta, Adagrad, Adamax, and AdamW), learning rates ($10^{-6}$ to $10^{-2}$), and weight decays ($10^{-5}$ to $10^{-2}$), with a fixed batch size of 256. Potential MLP models were chosen based on both accuracy (RMSE value) and training stability to ensure effective transfer learning to different datasets.

**2.3 Transfer learning interatomic potential**

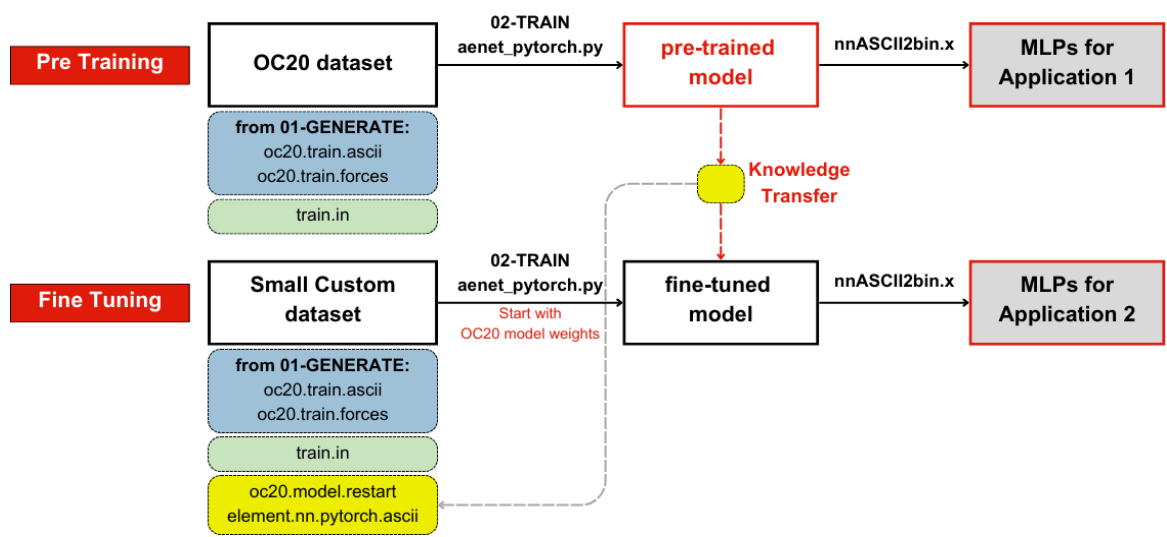

**Figure 3.** Technical workflow of transfer learning using ænet-PyTorch.

To train the MLP using the transfer learning approach, ænet-PyTorch requires the *model.restart* and *element.nn.ascii* from the source task, where the model was trained with a large and diverse database, in this case, the OC20 subsets. In the transfer learning process, the ænet-PyTorch framework initializes training on the target task, which involves a specific application database, with pre-trained weights instead of random initialized weights. This method anticipates that the knowledge gained from the source task can significantly enhance the accuracy and generalizability of MLPs to unseen data, particularly when the training database for the target task is very small. In this case, the architecture of ANN should be fixed during the process.

From the pre-trained MLP models constructed with the three OC20 subsets, we performed transfer learning to develop fine-tuned MLPs for our dataset of a CuAu alloy cluster with one adsorbed water molecule (CuAu/$H_2O$), using approximately 8,000 DFT reference structures from the *ab initio* molecular dynamics simulation (AIMD) trajectory (see Supporting Information section S1 for all AIMD details). Additionally, we tested the generalizability of the best MLP models for MD simulations of a larger CuAu alloy cluster with 6 water molecules (CuAu/6$H_2O$).

## 3. Results and Discussions

### 3.1 The selected OC20 subsets

Figures 4a, 4c, and 4e present the 2D histogram distribution of energy and forces for each OC20 subset in Table 1, overlaid with the original OC20 dataset distribution. Figures 4b, 4d, and 4f show the distribution of the number of atoms per structure for each subset. Additionally, Figures S3b, S3c, and S3d illustrate the element distributions for the elements included in each subset. The original dataset contains 56 elements (Figure S3a), with energy and force values concentrated around -4 eV/atom and 0 eV/Å, respectively. The dataset includes structures with atom counts ranging from about 2 to over 200 atoms per structure, with most samples containing 50 to 100 atoms per structure.

For the random subset, we aimed to randomly select 10% of the OC20 dataset to closely represent the original dataset. However, balancing the included elements, the energy-force distribution, and the number of atoms per structure proved challenging. As a result, we prioritized the distribution of energy over forces, given that force data is three times more prevalent. Ultimately, the random subset contains 56 elements with a distribution similar to the original dataset, depicting energy and force ranges from 0 to -12 eV/atom and -30 to 30 eV/Å, respectively.

The next two subsets were selected based on structures containing "Cu" and/or "Au". We were not concerned about the adsorbates since hydrogen and oxygen are dominant within the database (Figure S3a). First, we filtered structures that contain "Cu", "Au", and "Cu" + "Au", resulting in the CuAu-54 subset, which contains approximately 170,000 structures and 54 elements (Figure S3c). Compared to the random subset, the CuAu-54 subset has a narrower energy distribution but a wider range of force distribution.

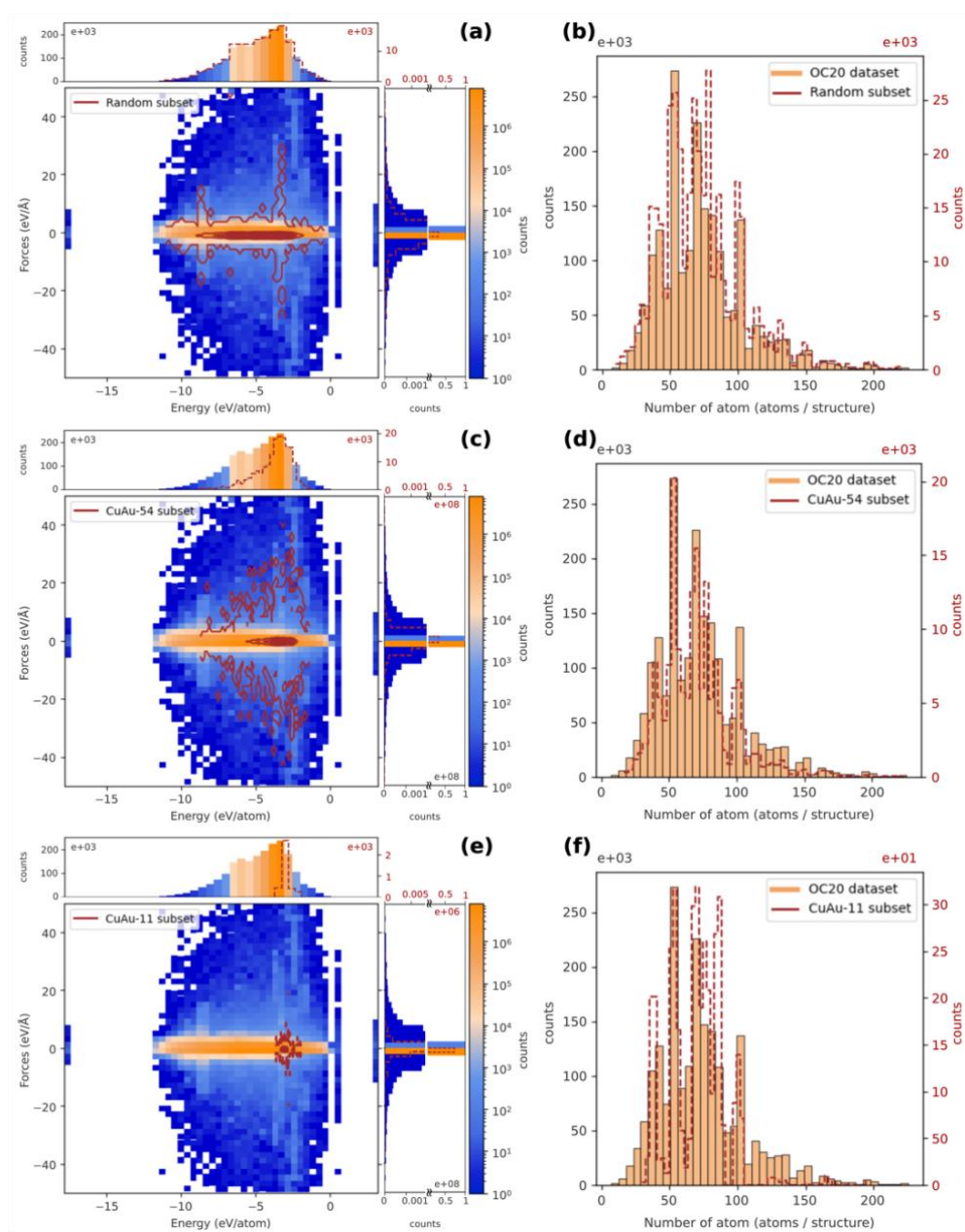

**Figure 4.** The energy-force 2D histogram of the OC20 dataset overlaid by the (a) random, (c) CuAu-54, and (e) CuAu-11 subsets. The distribution of the number of atoms per structure of the OC20 dataset overlaid by the (b) random, (d) CuAu-54, and (f) CuAu-11 subset.

The final subset, named CuAu-11, was selected based on its similarity to our target material CuAu/H$_2$O. This subset contains structures with only a combination of "Cu" + "Au" in a structure, resulting in 11 elements and a total of 4,000 structures. The energy and force values are concentrated around -3 eV/atom and 0 eV/Å, respectively. The structures in the CuAu-11 subset range from 20 to 100 atoms per structure.

### 3.2 Construction of the training sets

#### 3.2.1 The descriptor dimension of the training dataset

Varying the descriptor dimension is crucial for optimizing computational cost, feature representation, efficiency, and generalization[13,17,18]. This experiment aims to find the "sweet spot" for the model's effectiveness in predicting the properties of new atomic environments. Figure 5 shows the RMSE of the MLP model as a function of descriptor size, measured from independent models from the given subsets. The RMSE decreases significantly as the descriptor size increases from 36 to 44, indicating improved predictive accuracy. Beyond 44, further improvements are marginal. Thus, we selected a descriptor size 44 as the optimal balance between computational efficiency and predictive performance. These results align with previous findings on computational tractability and multi-species interaction capture[13]. This descriptor size adequately characterizes the local atomic environment, ensuring accurate MLP predictions without unnecessary computational costs.

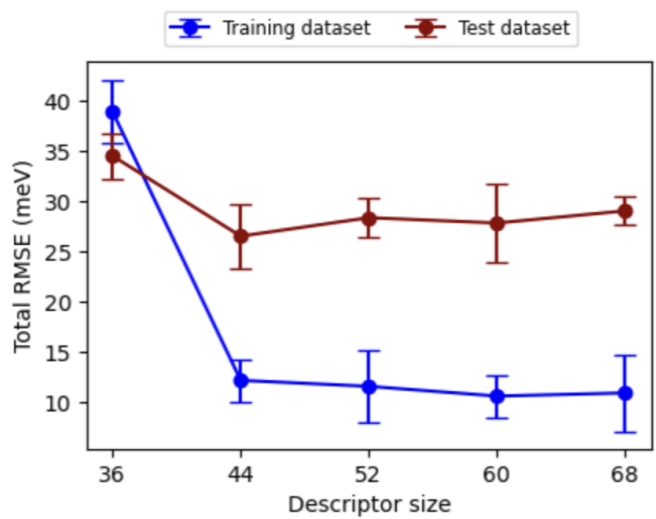

**Figure 5.** Accuracy of MLPs of the given subsets as a function of the descriptor dimension that represents the local atomic environment.

*3.2.2  The cutoff radius dimension of the training dataset*

The setup of radial and angular cutoffs significantly impacts the computational efficiency and accuracy of simulations. Learning from various practices using different descriptors that contain RDF and ADF components, the cutoff radius is set based on the type of interaction suitable for specific systems [8,19,20]. The radial cutoff determines the maximum distance within which atomic interactions are considered, which is essential for capturing relevant interactions without unnecessary computations that could slow down the simulation. For short-range interactions (typically 6-10 Å), this cutoff ensures computational efficiency and accuracy for small molecules and simple systems where long-range effects are minimal. Medium-range interactions (typically 10-12 Å) offer a balance between accuracy and computational cost, making them suitable for a broad range of materials science applications. For long-range interactions (typically 12-15 Å), the radial cutoff is extended to capture significant interactions in large biomolecules and complex materials, ensuring the model's accuracy in these scenarios.

The angular cutoff defines the level of detail in the angular dependencies of atomic interactions. Low-order angular cutoffs (typically 3-4) are appropriate for simple systems where detailed angular dependencies are not critical, providing basic interaction information with minimal computational cost. Medium-order cutoffs (typically 4-6) strike a balance for most materials science and chemistry applications, capturing essential angular information without excessive computational overhead. High-order angular cutoffs (typically 6-8) are vital for highly complex systems, providing detailed angular interaction information necessary for accurately modelling anisotropic materials and surface science studies.

Here, we experiment with the cutoff radius for low to medium-range applications using the given training dataset. The results in Figure 6 illustrate that a radial cutoff of 8 Å and an angular cutoff of 5 is optimal, considering both computational efficiency and accuracy for the given OC20 subsets.

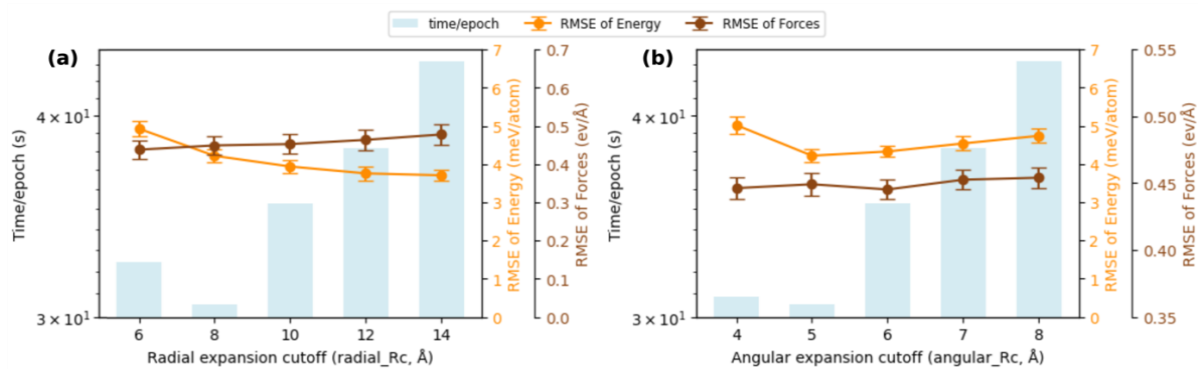

**Figure 6.** Accuracy of MLPs as a function of the descriptor dimension that represents the local atomic environment.

## 3.3 Pre-training interatomic potential based on the OC20 subsets

### *3.3.1 Training of energy*

While constructing the pre-training models, we observed that large learning rates (LR) often led to an unstable training process across most optimizers (Figures S6-S10). Optimizers such as Adam, Adadelta, and Adamax exhibited slow reductions in RMSE for both energy and force predictions, even when higher learning rates were applied. These Adam-based optimizers also showed varying reactions to changes in weight decay. Notably, AdamW and Adagrad were more efficient in handling both learning rate and weight decay simultaneously, providing better stability and performance in the training process[21]. Inspecting the learning processes is necessary, particularly for transfer learning purposes, because it ensures that the pre-trained model adapts well to the new, specific dataset. This inspection helps identify optimal hyperparameter settings that stabilize the training process and improve model performance. The careful monitoring can also reveal how well the model retains helpful knowledge from the pre-trained phase while adjusting to new data, ultimately enhancing the accuracy and reliability of the transfer learning application.

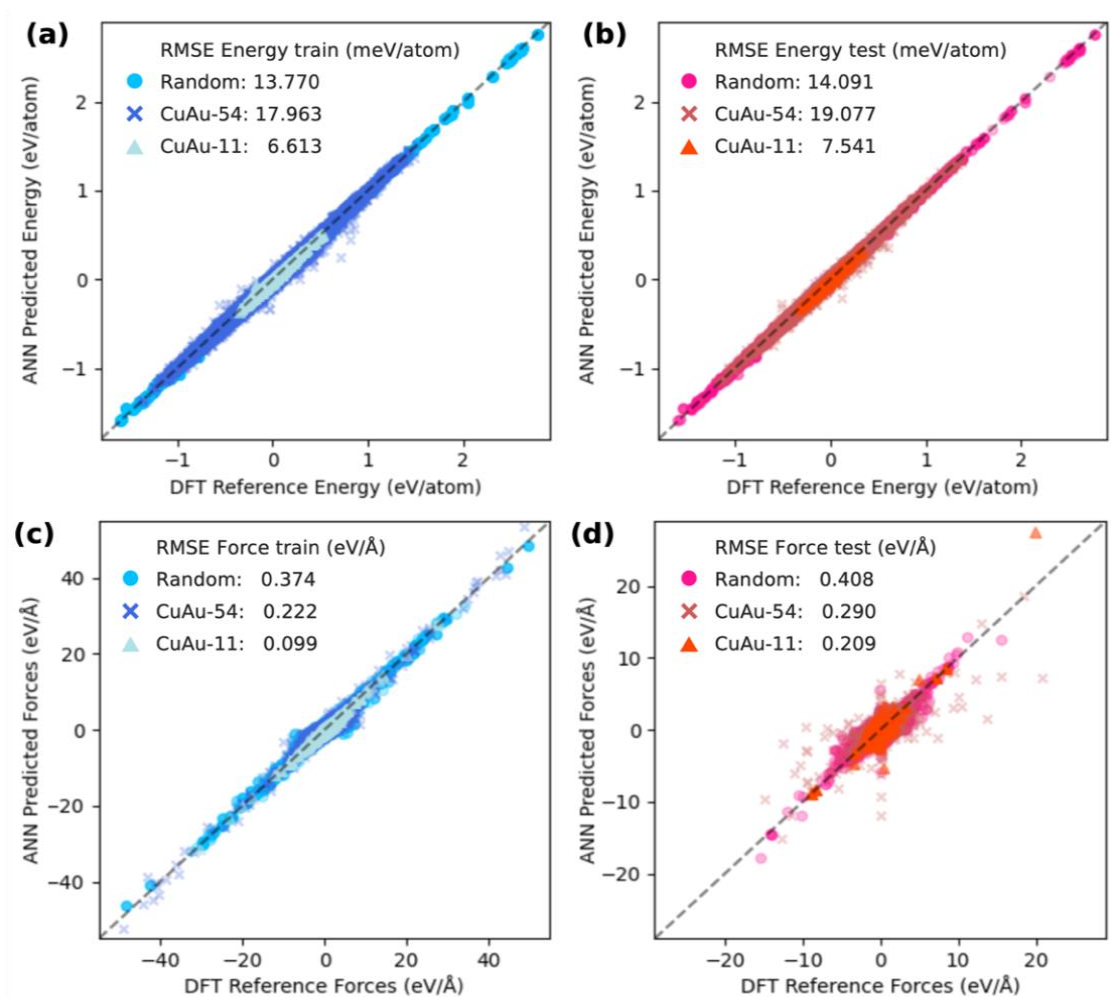

**Figure 7.** Evaluation of pre-training MLP models' prediction for each OC20 subset: Random, CuAu-54, and CuAu-11. (a) and (b) show the energy predictions, while (c) and (d) illustrate the forces predictions. Blue markers represent training data, and red markers represent test data.

Figure 7 displays the energy and force prediction power of the best MLP fitting results for each OC20 subset. Figures 7a and 7b show the energy predictions, while 7c and 7d illustrate the force predictions. In the energy prediction (Figure 7a and 7b), the results indicate that the CuAu-11 subset has the lowest RMSE for training and test data, suggesting it has the most accurate energy predictions among the subsets. For the training data (Figure 7a), the Random subset achieves an RMSE of 13.770 meV/atom, the CuAu-54 subset an RMSE of 17.963 meV/atom, and the CuAu-11 subset an RMSE of 6.613 meV/atom. For the test data (Figure 7b), the Random subset achieves an RMSE of 14.091 meV/atom, the CuAu-54 subset an RMSE of 19.077 meV/atom, and the CuAu-11 subset an RMSE of 7.541 meV/atom. We expect that these results arise from the nature of energy and force distribution within the subset.

*3.3.2 Training of forces*

Prior research indicates that including at least 10% of forces in the training set is adequate[16]. In the original article of aenet-PyTorch, López-Zorilla et al. discuss the balance between energy and force information in training data[16]. It is emphasized that while all structures contribute to energy fitting, only a subset is used for force training. An experiment removing energy-only structures from training showed a disproportionate impact on energy prediction accuracy, mainly when data coverage is sparse. Force predictions were less affected. This implies that while force data enhances the model around known structures, predictions for configurations far from the training set are less reliable.

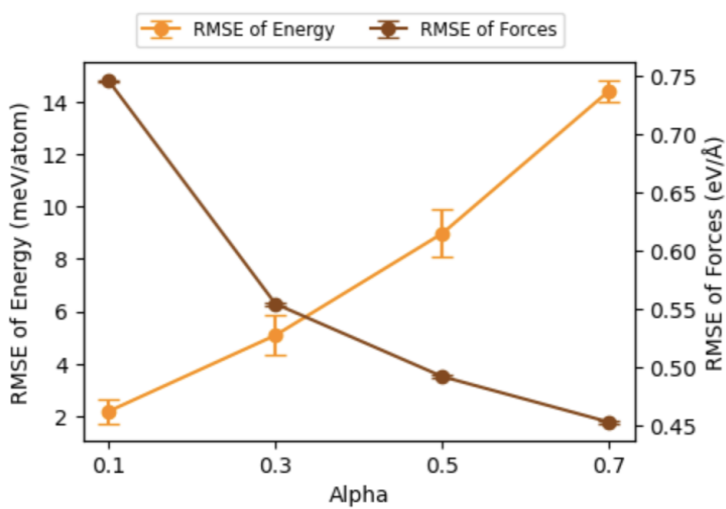

**Figure 8.** The accuracy of energy and forces prediction as the function of alpha (α).

Besides the hyperparameter mentioned in the energy training section, one hyperparameter that greatly influences the force prediction accuracy is alpha (α). In ænet-PyTorch, the parameter α in the combined loss function balances energy and force components during training[16]. As α increases, the model prioritizes minimizing force loss, leading to better force prediction accuracy but worse energy prediction accuracy. This trade-off occurs because higher α shifts the model's focus and capacity towards fitting forces more precisely, potentially at the expense of energy predictions. Our results in Figure 8 and Figure S5 show that while increasing α improves force RMSE, it simultaneously raises the energy RMSE, indicating that the model's ability to predict energy diminishes as it prioritizes force accuracy.

Further, in the previous conclusion based on the $TiO_2$ database[16], the α values of 0.1~0.3 can be optimal for general application. However, we found that α = 0.1 leads to a loose force prediction in our training dataset (Figure S5). The α of 0.3~0.5 might be suitable for the OC20 subsets. This shift can be attributed to the increased complexity and diversity of the OC20 dataset, which includes a wider variety of atomic environments and interactions. In such a dataset, a higher α is necessary to ensure the model allocates sufficient capacity to capture the more nuanced and varied force interactions accurately. As a result, the model can better generalize across the diverse configurations present in OC20, leading to improved overall performance, especially in energy predictions. This indicates that the balance between energy and force accuracy, governed by α, is dataset-dependent and needs to be carefully tuned to match the specific requirements and characteristics of the training data.

Overall, similar to the energy predictions, the CuAu-11 subset demonstrates the lowest RMSE for both training and test data in force predictions. However, the MLP model from the CuAu-54 subset has better predictability than the MLP model from the Random subset. For the training data (Figure 7c), the Random subset achieves an RMSE of 0.374 eV/Å, the CuAu-54 subset an RMSE of 0.222 eV/Å, and the CuAu-11 subset an RMSE of 0.099 eV/Å. For the test data (Figure 7d), the Random subset achieves an RMSE of 0.408 eV/Å, the CuAu-54 subset an RMSE of 0.290 eV/Å, and the CuAu-11 subset an RMSE of 0.209 eV/Å.

### 3.4 Transfer learning interatomic potential to CuAu/H$_2$O dataset

*3.4.1 Trial on the direct application*

Before applying transfer learning, we tested our hypothesis by evaluating MLPs trained on OC20 (our pre-trained models) on two independent test sets to predict adsorption energies: Pd-Au/H$_2$/O$_2$[22] and Cu-Au-Pt/O$_2$. Both systems exhibited significant discrepancies between the adsorption energies predicted by the pre-trained MLP models and the reference DFT values.

The results from these tests are detailed in the Supplementary Information sections S2–S4. As expected, these tests confirmed that the pre-trained models cannot be used directly for our complex catalyst models.

*3.4.2   Accuracy comparison of transfer learning from different subsets*

Herein, we highlight the benefits of using the OC20 pre-trained model for transfer learning. Figure 9 shows the energy and force prediction accuracy of the MLPs for CuAu/$H_2O$, constructed from scratch and using transfer learning from pre-trained MLPs on OC20 subsets. The subsets used for transfer learning include Random (TF-Random), CuAu-54 (TF-54), and CuAu-11 (TF-11). The performance is measured in terms of the RMSE for both energy and forces. The bar plot illustrates the RMSE values for both training ($E_{train}$ and $F_{train}$) and test ($E_{test}$ and $F_{test}$) datasets.

For the energy predictions, the TF-11 subset exhibits the lowest RMSE for both the training and test datasets, followed by the TF-54 and TF-Random subsets. This trend indicates that the CuAu-11 subset, which is closely related to the CuAu/$H_2O$ system, provides the most significant improvement in prediction accuracy. The energy RMSE for the model trained from scratch is higher compared to all the transfer learning models, highlighting the effectiveness of using pre-trained models.

Similarly, for the force predictions, the TF-11 subset again shows the lowest RMSE values for both training and test datasets, followed by the TF-54 and TF-Random subsets. The force RMSE for the model trained from scratch is the highest, further demonstrating the benefit of transfer learning in improving the prediction accuracy of MLPs.

Overall, these results clearly demonstrate that transfer learning from OC20 pre-trained models significantly enhances the accuracy of energy and force predictions for the CuAu/$H_2O$ system. Among the subsets, the CuAu-11 subset offers the greatest benefit, likely due to its specific relevance to the target system. This result underscores the importance of selecting appropriate pre-trained datasets for transfer learning to achieve optimal performance in machine learning potentials.

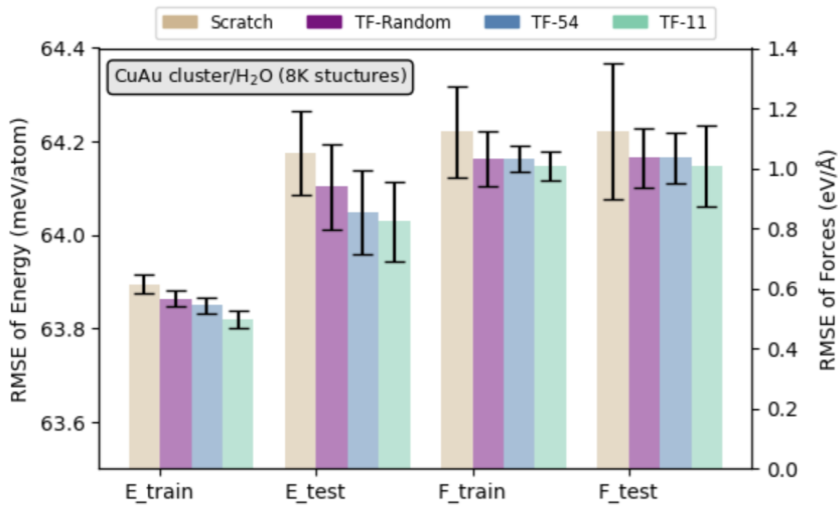

**Figure 9.** Energy and force prediction accuracy of MLPs for CuAu/$H_2O$ constructed from scratch and using transfer learning from OC20 subsets: Random (TF-Random), CuAu-54 (TF-54), and CuAu-11 (TF-11).

### 3.4.3 Application for MD simulation

The superior generalizability of the transfer learning (TF) model is evident through molecular dynamics (MD) simulations. Here, we used 574 structures of CuAu/$H_2O$ instead of all 8,000 structures to construct the MLPs. We then compared the performance of MLPs derived from different training approaches: training from scratch (S model) and transfer learning (TF model). The MLP training results, shown in Table 1, demonstrate that both MLP models achieved the same level of accuracy.

Table 2. Comparison of the RMSE in MLP training using different approaches on the CuAu/$H_2O$ dataset.

| Approach | $E_{train}$ (meV) | $E_{test}$ (meV) | $F_{train}$ (meV) | $F_{test}$ (meV) |
|---|---|---|---|---|
| Direct learning on CuAu/$H_2O$ dataset (**S model**) | 86.373 | 92.709 | 109.107 | 114.361 |
| Transfer learning from OC20 → CuAu/$H_2O$ (**TF model**) | 86.312 | 92.449 | 112.721 | 115.499 |

Further, to assess the extrapolation performance, we conducted 250 ps MLP-MD simulations of a CuAu/6$H_2O$ system using both MLPs. The simulations were performed under the canonical ensemble (NVT) using a Bussi thermostat with timesteps of 0.5 fs at the temperature of 300 K. The radial distribution function (RDF) derived from the S model's trajectory (Figure 10a) exhibits abnormal Au-O distances, with the RDF of Au-O peaking at 1.065 Å. In contrast, the TF model shows a peak at 2.195 Å. Both MLPs indicate Cu-O distances for adsorbed water in the range of ~2.1-2.2 Å. However, the S model's simulation reveals some abnormal Cu-O distances of less than 1.5 Å, a phenomenon not observed with the TF model. These results support that the prior knowledge from the OC20 pre-trained model is conserved and effectively transferred to enhance our specific task, even though this improvement is not reflected by RMSE in MLP training.

Moreover, we noted that the OC20 pre-trained model cannot be directly applied to run MD simulations of CuAu/6H$_2$O (Figure S4) due to the absence of metal cluster data in the OC20 database. Therefore, transfer learning becomes crucial for leveraging large chemical databases for specific chemical systems and applications of interest.

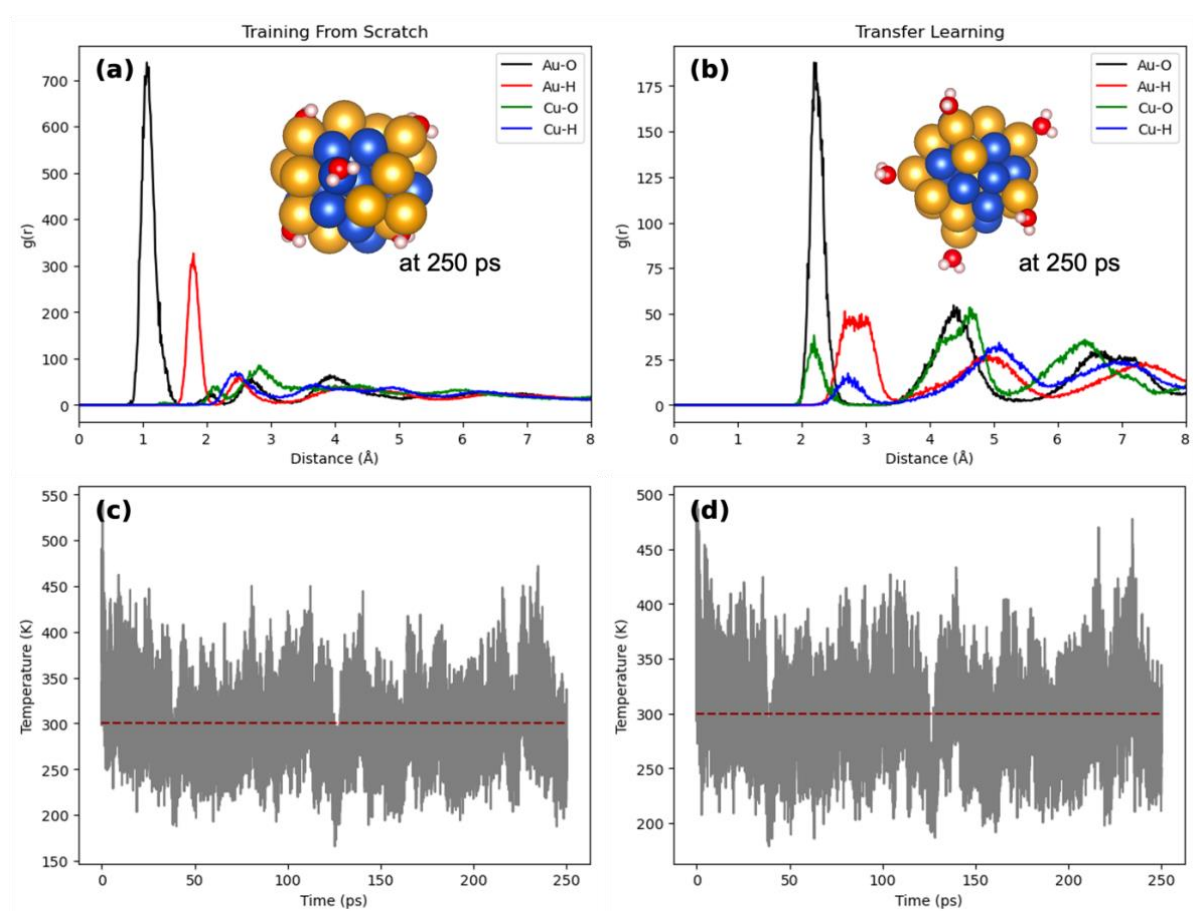

**Figure 10.** (a) and (b) the radial distribution function (RDF) and MD snapshot at 250 ps. (c) and (d) temperature during MD simulations derived from S and TF models, respectively.

*3.4.4   Influence of fine-tuning data size in transfer learning accuracy*

The result above shows that transfer learning can enhance MLP stability during the MD application, even though the RMSE did not indicate significant improvement. Thus, we assess the effect of the fine-tuning dataset size on the prediction accuracy of transfer learning. This is achieved by comparing the RMSE of the TF model to the RMSE of the S model. Given that the RMSE scale differs for the two dataset sizes (8K and 0.6K), it is crucial to use a paired t-test to evaluate the statistical significance of the differences between TF and S results. The t-test p-value provides a robust measure of whether the observed differences are likely due to random chance or if they indicate a genuine effect of the dataset size on prediction accuracy[23].

The visualized p-values in Figure 11 indicate the impact of data size on the prediction accuracy of energy and forces when comparing the TF model and S model. For Energy, the p-values for both the 8K (0.239) and 0.6K (0.353) data sizes are above the significance threshold of 0.05[23]. This suggests that the differences in prediction accuracy between the TF and S methods for energy are not significant, regardless of the data size. Thus, data size does not remarkably appear to affect the prediction accuracy for energy.

In contrast, the p-values for the force component show a different trend. The p-value for the 8K data size (0.010) is below the significance threshold, indicating a statistically significant difference in force prediction accuracy between the TF and S methods. However, the p-value for the 0.6K data size (0.306) is above the threshold, suggesting no significant difference in prediction accuracy at this smaller data size. This implies that a larger data size (8K) is necessary to detect significant differences in the prediction accuracy of forces between the TF and S conditions, highlighting the impact of data size on the reliability of force predictions.

These results indicate that transfer learning's benefits become more pronounced with larger datasets. While the RMSE values for the smaller dataset do not show a significant advantage for the TF model, the MD simulations reveal its superior stability and physical accuracy. This highlights that RMSE alone does not always indicate model performance, especially in dynamic simulations. Transfer learning, therefore, proves crucial for leveraging large chemical databases to enhance specific tasks, ensuring better generalization and stability in practical applications.

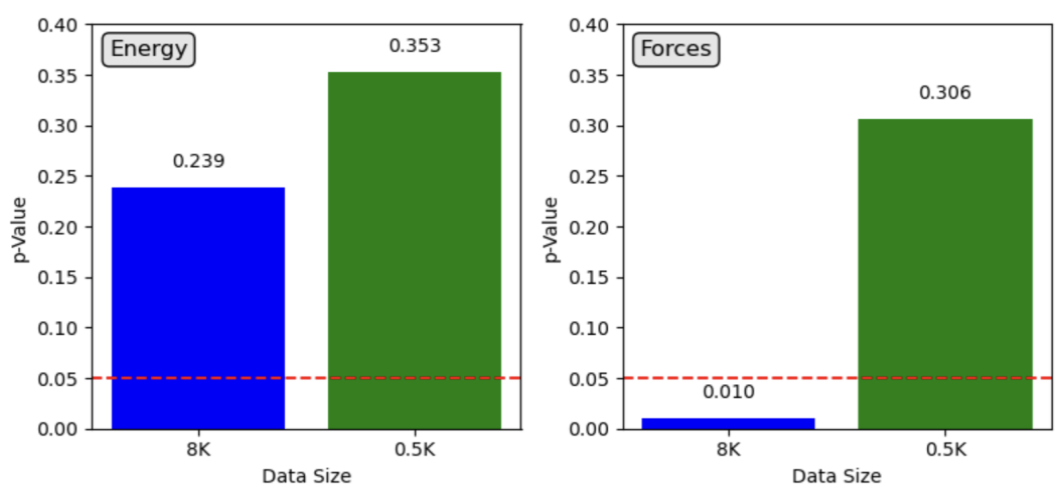

**Figure 11.** p-Values for (a) energy and (b) force prediction accuracy across different fine-tuning data sizes of CuAu/$H_2O$.

## 4. Conclusions

We aimed to explore a cost-effective approach for constructing machine learning potentials (MLPs) for challenging catalyst-adsorbate systems using transfer learning. By leveraging an available public dataset and carefully selecting appropriate subsets, we demonstrated two strategies: random selection of subsets and specific selection of subsets. Our findings indicate that specific subset selection, based on chemical environment similarity, results in better accuracy, underscoring the importance of dataset relevance in enhancing MLP transfer learning. We further investigated the feasibility of transferring the MLP model to a smaller fine-tuning dataset with 600 data points. While the RMSE values for energy and force predictions did not show significant improvement, the generalizability of the transfer learning MLP was evident through stable MD simulation results. The MLP derived from transfer learning proved to be more stable and accurate for long-term MD simulations, whereas the MLP constructed from scratch did not achieve comparable stability. Additionally, our results highlight the advantage of using a larger fine-tuning reference dataset. A larger dataset significantly enhances the performance and applicability of the MLP, making it more robust for practical applications. Furthermore, we learned that evaluating MLPs based solely on RMSE values is insufficient. Applying the MLPs in practical simulations is crucial to ensure their usefulness and effectiveness. Overall, this study underscores the effectiveness of transfer learning in constructing accurate and stable MLPs for complex material systems, emphasizing the importance of dataset selection, size, and practical application in optimizing model performance.

## AUTHOR INFORMATION

Corresponding Author
Nongnuch Artrith
Materials Chemistry and Catalysis, Debye Institute for Nanomaterials Science, Utrecht University
Universiteitweg 99, 3584 CG Utrecht, The Netherlands
Tel: +31 (0)30 253 1600
e-mail: n.artrith@uu.nl

**Data Availability Statement**

The authors declare no competing financial interest. The data from the Density Functional Theory (DFT) calculations can be obtained from the GitHub repository at https://github.com/atomisticnet/XXXXX

**Acknowledgements**

The numerical calculations were carried out on the TSUBAME4.0 supercomputer at the Tokyo Institute of Technology, supported by the MEXT Project of the Tokyo Tech Academy for Convergence of Materials and Informatics (TAC-MI). We thank the Dutch National e-Infrastructure and the SURF Cooperative for providing the computational resources used in the DFT calculations, ænet-PyTorch training, and the development of the transfer learning model. This work was funded by a start-up grant (Dutch Sector Plan) from Utrecht University awarded to N.A.

# Supporting Information

# A cost-effective strategy of enhancing machine learning potentials by transfer learning from a multicomponent dataset on ænet-PyTorch

*An Niza El Aisnada[a,b], Kajjana Boonpalit[b,e], Robin van der Kruit[b], Koen M. Draijer[b], Jon Lopez-Zorrilla[c], Masahiro Miyauchi[a], Akira Yamaguchi[a,d], Nongnuch Artrith[b*]*

[a] Department of Materials Science and Engineering, School of Materials and Chemical Technology, Tokyo Institute of Technology, 2−12−1 Ookayama, Meguro-ku, Tokyo 152−8552, Japan

[b] Materials Chemistry and Catalysis, Debye Institute for Nanomaterials Science, Utrecht University, 3584 CG Utrecht, The Netherlands

[c] Physics Department, University of the Basque Country (UPV/EHU), Leioa, Basque Country, Leioa, Spain

[d] Biofunctional Catalyst Research Team, RIKEN Center for Sustainable Resource Science, 2-1 Hirosawa, Wako, Saitama 351-0198, Japan

[e] School of Information Science and Technology, Vidyasirimedhi Institute of Science and Technology, Rayong, Thailand

## S1. CuAu/H$_2$O dataset

All *ab initio* molecular dynamics (AIMD) simulations were run for a total time of 10 ps (picoseconds) with a time step of 1 fs to obtain converged trajectories for CuAu nanoparticles with 2, 4, 6, and 8 water (H$_2$O) molecules. The FHI-aims code [1], developed at the Fritz Haber Institute in Berlin, was used for the AIMD calculations. The generalized gradient approximation (GGA) exchange-correlation functional by Perdew, Burke, and Ernzerhof (PBE) [2] was employed in combination with relativistic corrections at the level of the Atomic ZORA approximation [3], using a time step of 1 fs. The MD runs were conducted with the NVT canonical ensemble using the Bussi-Donadio-Parrinello thermostat at 400 K. Additionally, the method by Tkatchenko and Scheffler [4] was used to correct for the missing van der Waals interactions at this level of DFT. For optimal computational efficiency, AIMD simulations with FHI-aims employed the predefined light basis set (4th order expansion of the Hartree potential, radial integration grids with 302 points in the outer shell, and a tier 1 basis set).

## S2. Cu-Au-Pt/O$_2$ dataset

For the calculations involved in the Pt addition on an Au-terminated 100 CuAu surface and the subsequent O$_2$ dissociation on the Pt-coated surface, DFT calculations with FHI-aims code [1], were utilized. All calculations were performed with the PBE functional [2], atomic ZORA scalar relativistic effects [3] and a SCF convergence of 1E-6 e/a_0^3.

Filling of the Au-terminated surface with Pt atoms followed three different addition schemes: sequentially, as a cluster, or as a monolayer. For each addition step, a two-step geometry optimization was performed with the 'light' and 'tight' basis sets and the Tkatchenko-Scheffler dispersion model [4]. The geometry was optimized with the BFGS algorithm and a convergence of 5E$^{-3}$ eV/AA on a 12 x 12 x 1 k-grid. One of the obtained Pt-coated surfaces was used for the dissociation investigation of O$_2$ on the Pt-coated surface using a nudged elastic band (NEB) as implemented in ASE, where the 'light' basis set was used without dispersion correction on a 5 x 5 x 1 k-grid. In an initial run, 5 images were used between the beginning and end minima (O$_2$ and 2 O on the surface), followed by Climbing Image NEB between the structures around the observed maximum to obtain the activation energy.

## S3. Direct application of pre-training MLP model on Cu-Au-Pt/O$_2$

We calculated the adsorption energies of the Au-terminated CuAu (100) surface during Pt addition, followed by predicting the relative energy during O$_2$ dissociation on the Pt-coated surface (DFT at PBE level with FHI-aims package [1], from Robin van der Kurit's Master Thesis 2023 [5]). The average predictions for the MLPs during addition and dissociation are shown in Figure S1 (a) and (b). The MLP energy increases, whereas the DFT results show a decrease in adsorption energy. Additionally, the relative NEB energy shows a slight decrease but lacks the distinct activation energy and significant energy drop expected for oxygen dissociation on a Pt-coated surface. In neither case do the MLP results follow the same trends as the DFT results, indicating that the models may be lacking in some aspects. Based on the dissociation results, it is possible that these discrepancies arise from the calculations being outside the training domain, causing the model to struggle with these structures. Another observation from the Pt addition results, when comparing the individual model outcomes, is that the models do not produce similar plot shapes in this scenario. This may indicate insufficient training data in the random selection or a difference emerging during training. The Pt-coated surfaces used in the dissociation part of this task do not have Pt atoms following the existing CuAu lattice. Instead, the surface features regions with higher (closer together) and lower Pt concentrations, resulting in a distorted minimum structure. However, this does not explain the poor performance for Pt addition, as the highly regular Pt-coated structure is part of the dataset.

We compared the results obtained with DFT to three different MLP models trained with ænet in two scenarios: the sequential addition of Pt on the Au-terminated surface (Figure S1a) and the nudged elastic band calculations for $O_2$ dissociation into 2 O atoms (Figure S1b). For the Pt addition, the MLP models indicated an increase in adsorption energy as the number of Pt atoms on the surface increased. However, this trend did not align with the DFT results. Similarly, for $O_2$ dissociation, two of the MLP models showed a decrease in relative energy, which is expected for $O_2$ dissociation on Pt, but they did not produce the relatively smooth plot shown by the DFT calculations. This discrepancy highlights the limitations of the MLP models in accurately capturing the adsorption and dissociation processes compared to DFT.

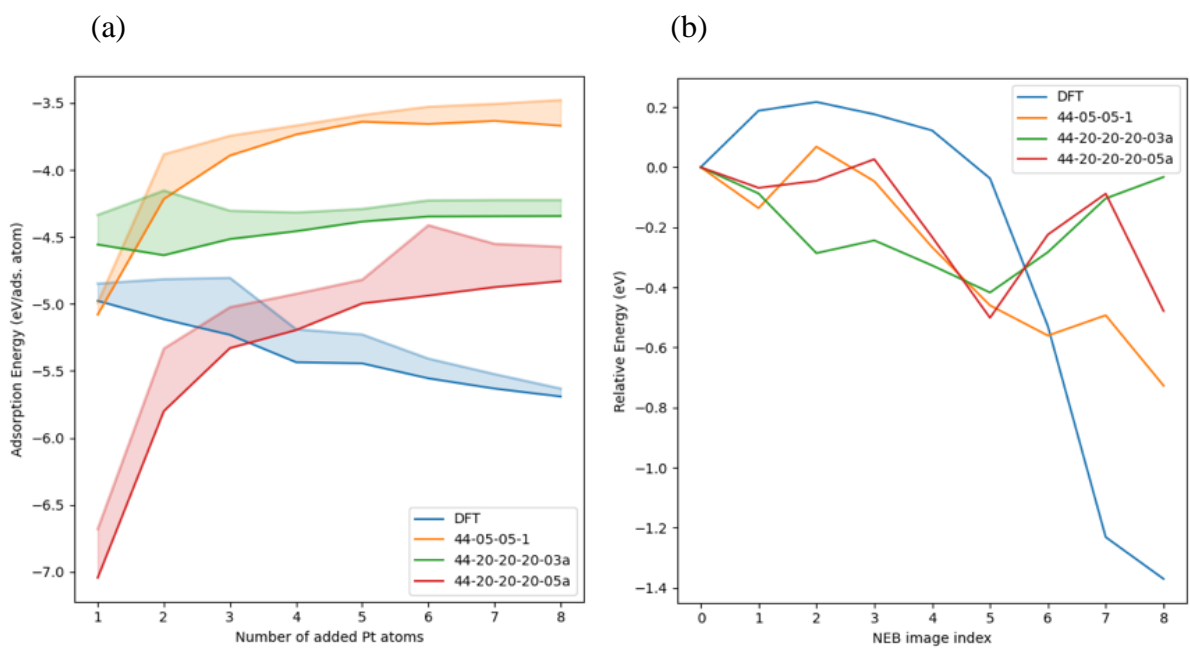

**Figure S1.** Comparison of three MLPs ænet-Pytorch models to the DFT results for **(a)** the sequential addition of Pt on an Au-terminated (100) CuAu surface where the top two layers are allowed to move. For each energy model, the lowest energies are shown with a solid line, while the shaded area represents the range of energies for a given number of Pt atoms. **(b)** the structures obtained during DFT nudged elastic band (NEB) calculations for the dissociation of $O_2$ on a Pt-coated surface.

## S3. Direct application of pre-training MLP model on H$_2$/O$_2$ adsorption on Pd-Au alloy

We calculated the adsorption energies of H$_2$ and O$_2$ on a Pd-Au alloy surface. The bottom five layers are a random mixture of Pd and Au in a 0.45:1 ratio (frozen during calculation), and the surface layer is made up out of 16 atoms which vary in composition of 0:1 Pd:Au to 1:0 Pd:Au. The adsorption energy was evaluated for the adsorbate for the same on top adsorption site in each model. This adsorption site was where the Pd atoms were introduced into the full Au surface layer. Computational details for the DFT calculation can be found in Refence [6]. We found that the 44-05-05-1-03a and 44-20-20-20-03a potentials have an absolute error in the O$_2$ adsorption energy smaller than 1 eV with the surface models where the Pd fraction is below 0.5 (Figure S2a). Interestingly, the error of all three potentials is smaller for a surface with a 1:0.07 Pd:Au ratio compared to a full Pd surface. Although the worst performing potential for the full O$_2$ adsorption dataset, the 44-20-20-20-05a potential has an error of only 0.04 eV for the full Au surface. The 44-05-05-1-03a potential has an mean absolute error (MAE) of 0.61 eV for the O$_2$ adsorption dataset, which is the lowest of the three tested machine learned potentials (Table S2). As the adsorption energies are in the range of 0.76-2.26 eV, this MAE is too large for the machine learning potentials to be reliable for the estimation of the energy of O$_2$ adsorption on Pd-Au alloy surfaces.

In the case of H$_2$ adsorption energies, the 44-20-20-20-03a and 44-20-20-20-05a potentials perform better than the 44-05-05-1-03a with the low Pd surface fractions, with an absolute error below 1 eV (Figure S2b). In the case of a pure Au surface, the 44-20-20-20-03a potential performs best with an absolute error of 0.02 eV. However, for the estimation of the H$_2$ adsorption on the Pd-rich surfaces, the 44-20-20-20-03a potential has absolute errors above 3 eV. In this regime, the 44-20-20-20-05a has an absolute error for the H$_2$ adsorption energy on the pure Pd surface of only 0.26 eV. From the tested potentials, the 44-20-20-20-05a potential has the lowest MAE of 0.51 eV for the H$_2$ adsorption dataset (Table S2). Similar to the O$_2$ adsorption energy dataset, the MAEs of the machine learning potentials make them unreliable for the prediction of H$_2$ adsorption energies on Pd-Au alloy surfaces, which had a range of 0.06-1.54 eV.

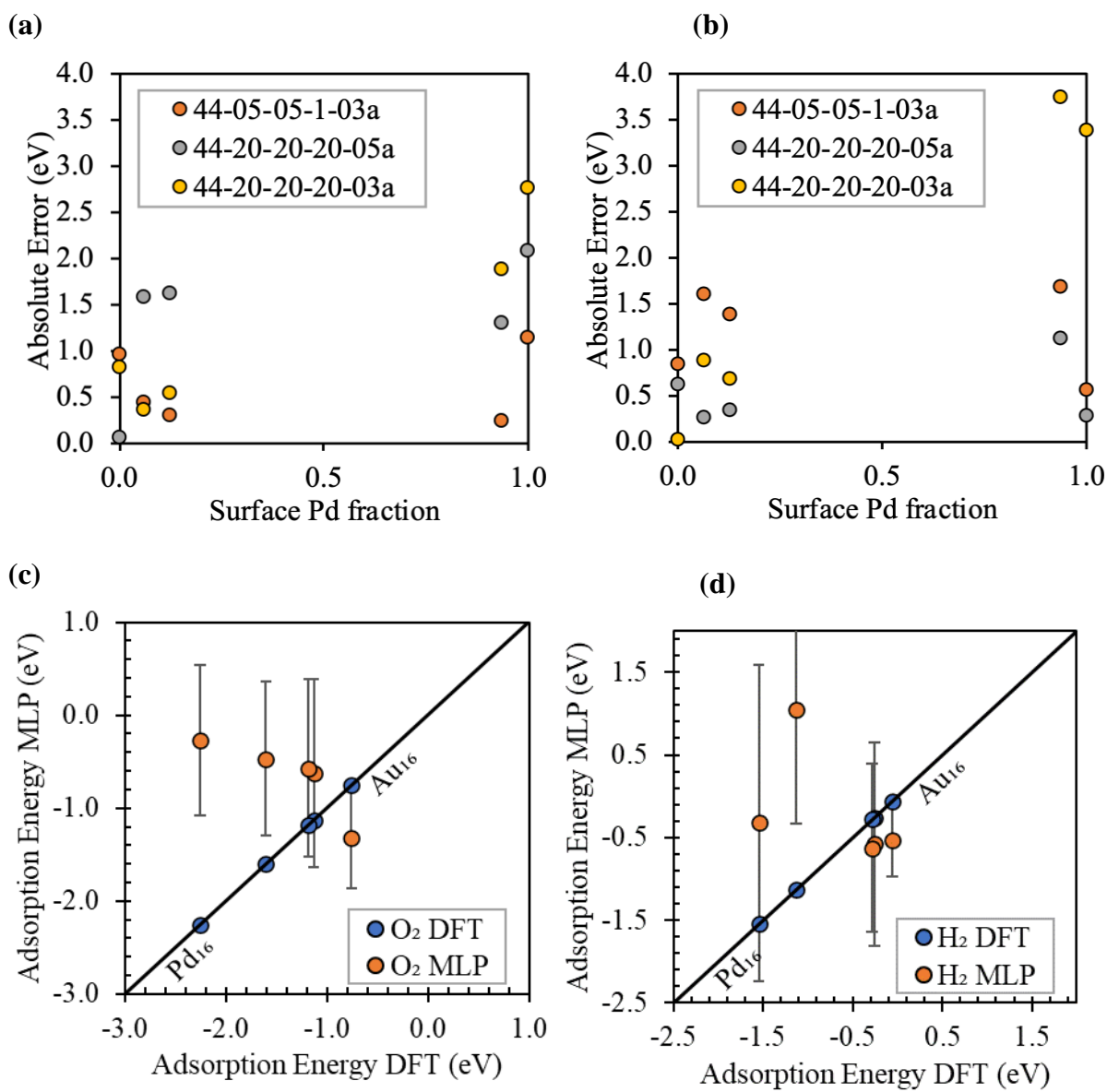

**Figure S2.** (a) Absolute error between the OC20 machine learning potentials and PBE+TS DFT for the calculated adsorption energy of molecular $O_2$ on random alloy Pd-Au (111) surfaces. (b) Absolute error between the OC20 machine learning potentials and PBE+TS DFT for the calculated adsorption energy of $H_2$ on random alloy Pd-Au (111) surfaces. (c-d) Average predictions from 3 MLPs for the adsorption energy of $O_2$ and $H_2$ on random alloy Pd-Au (111) surfaces.

**Table S2:** Mean Average Error (MAE) in eV per machine learning potential (MLP) for $O_2$/$H_2$ adsorption on Pd-Au alloy.

| MLP | MAE (eV) $O_2$ adsorption | MAE (eV) $H_2$ adsorption |
|---|---|---|
| **44-05-05-1-03a** | 0.61 | 1.20 |
| **44-20-20-20-05a** | 1.32 | 0.51 |
| **44-20-20-20-03a** | 1.26 | 1.73 |

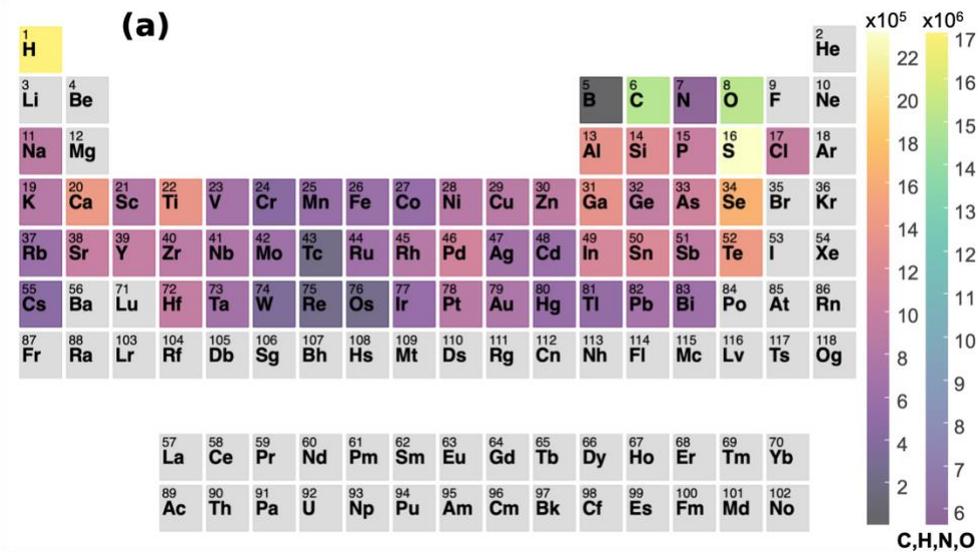

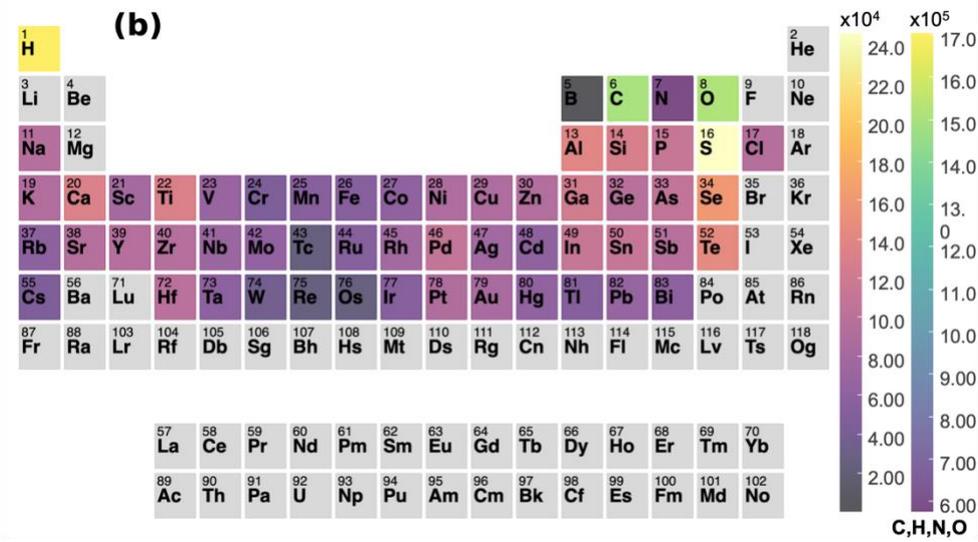

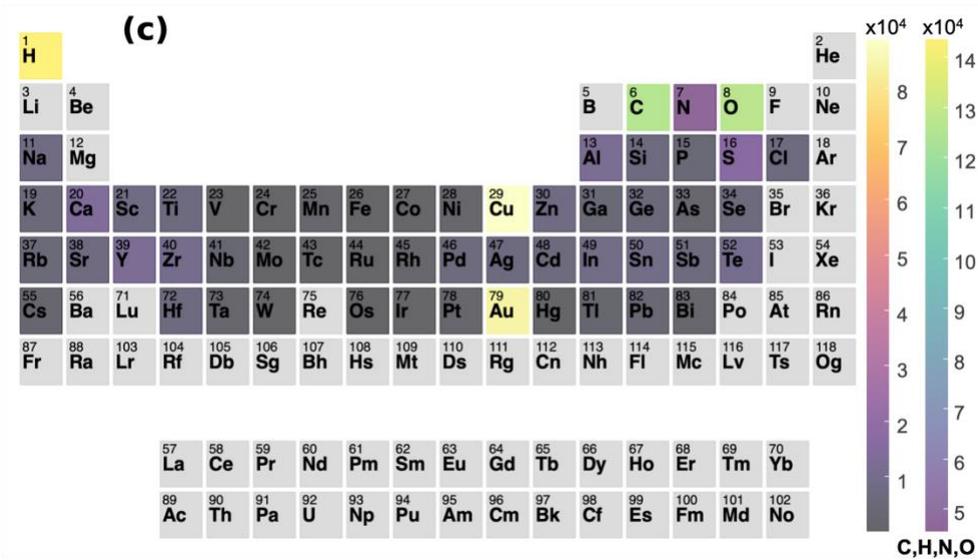

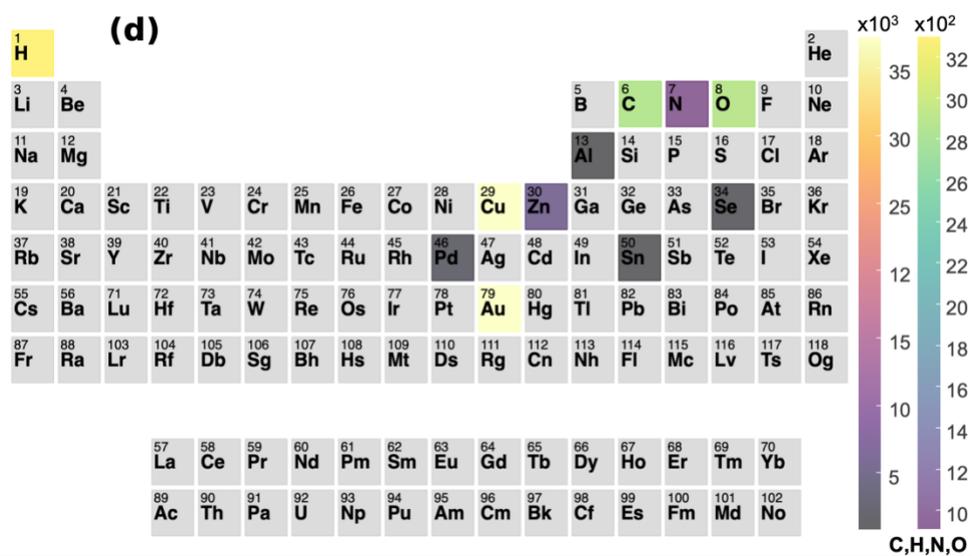

**Figure S3**. Elements distribution of the (a) OC20 database, (b) Random subset, (C) CuAu-54 subset, and (D) CuAu-11 subset.

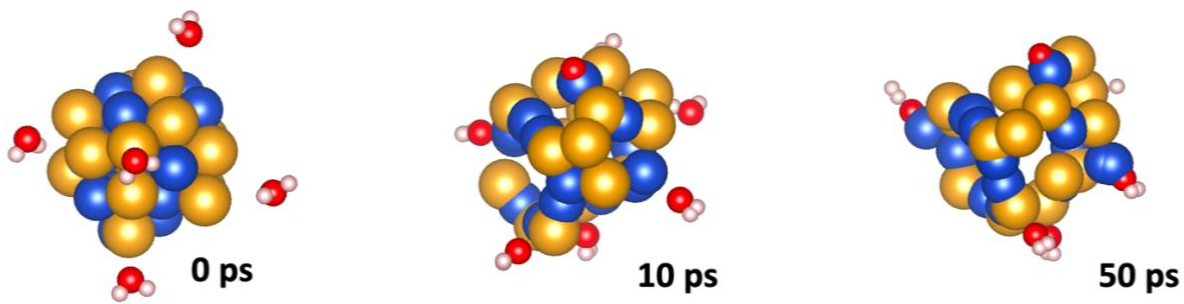

**Figure S4**. MD snapshots of CuAu/6H$_2$O at 0, 10, and 50 ps, performed by OC20 pre-trained model directly. The structure is distorted by inaccurate MLP due to the absence of metal cluster data in the OC20.

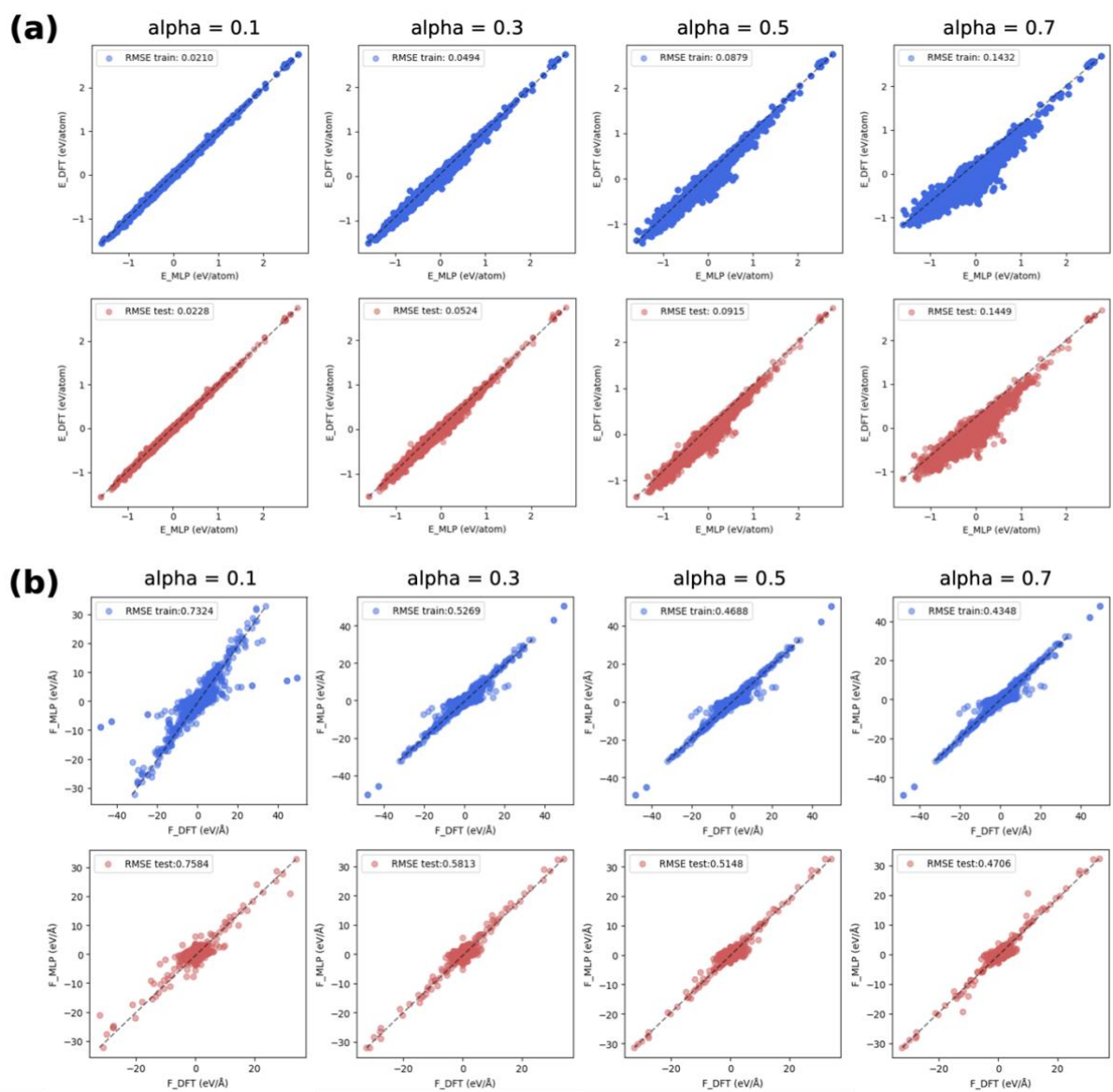

**Figure S5.** MLP accuracy according to the variation of alpha value ($\alpha$ = 0.1, 0.3, 0.5, 0.7) trained on the OC20 subset. The blue dots indicate the training data and red dots indicate test data during the training process.

## LR = 10⁻⁶

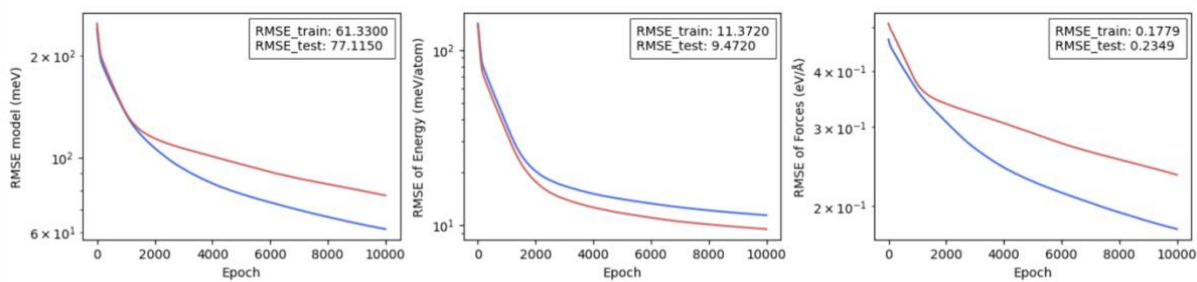

## LR = 10⁻⁵

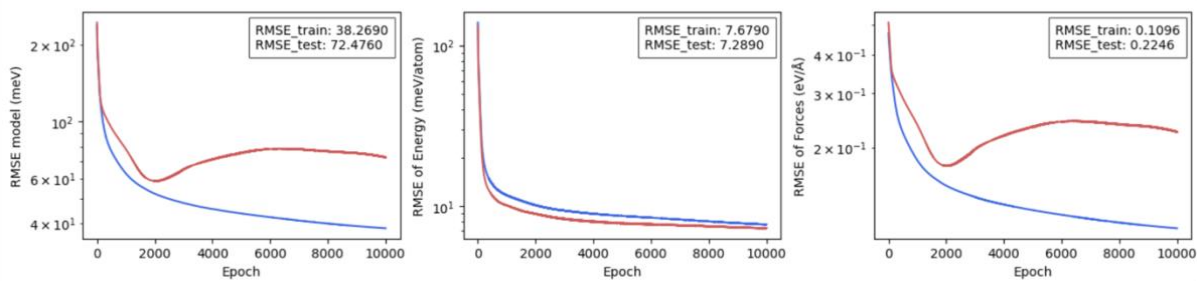

## LR = 10⁻⁴

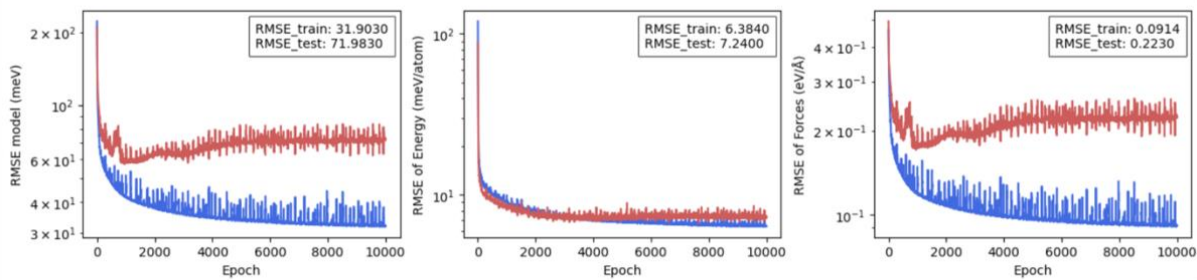

## LR = 10⁻³

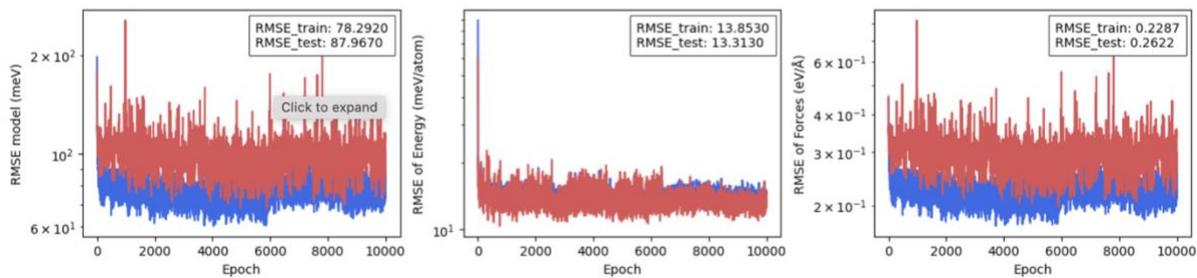

## LR = 10⁻²

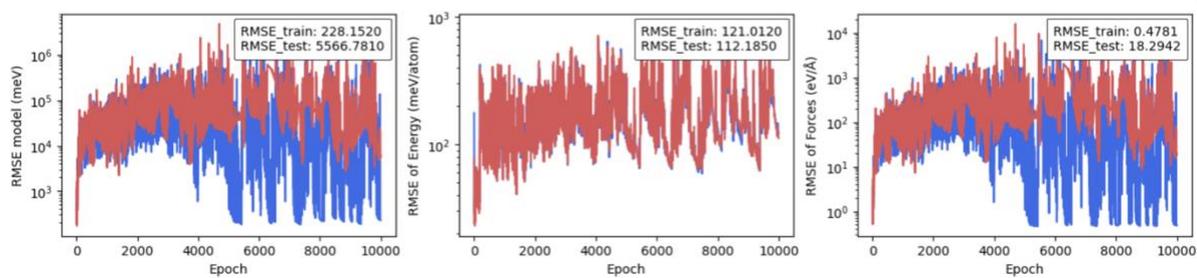

**Figure S6**. The overall prediction convergence (including energy and force) during the MLP training process for the OC20 subset using the Adam optimizer with LR from $10^{-2}$ to $10^{-6}$.

**LR = 10⁻⁶**

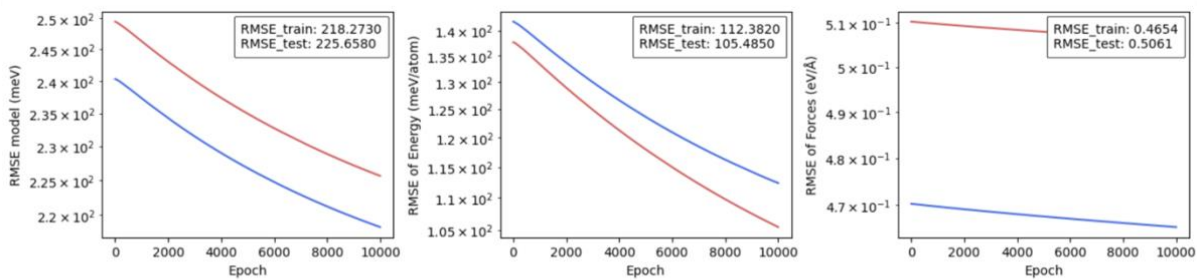

**LR = 10⁻⁵**

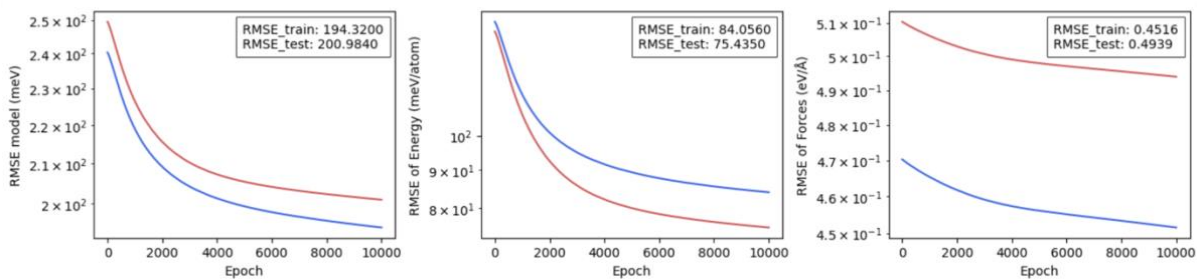

**LR = 10⁻⁴**

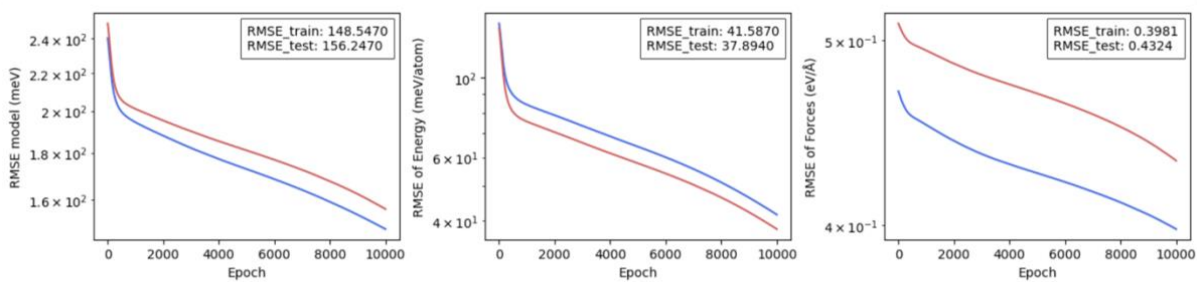

**LR = 10⁻³**

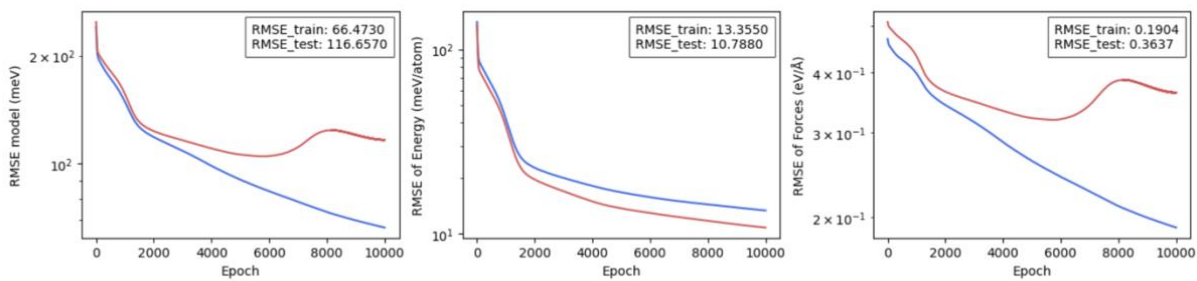

**LR = 10⁻²**

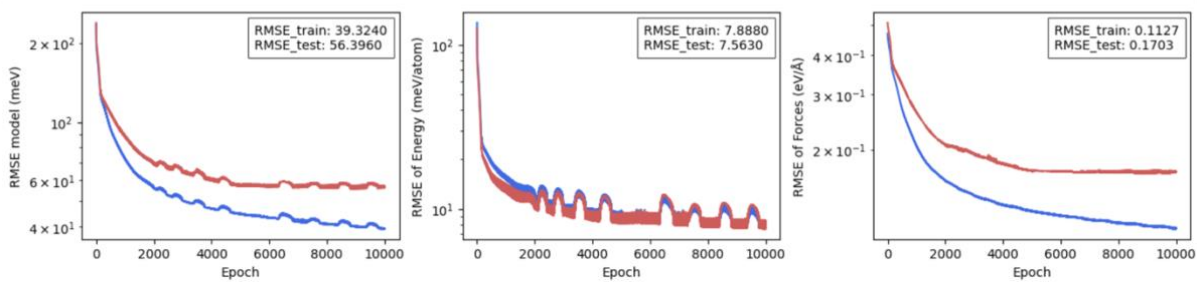

**Figure S7**. The overall prediction convergence (including energy and force) during the MLP training process for the OC20 subset using the Adadelta optimizer with LR from $10^{-2}$ to $10^{-6}$.

**LR = 10⁻⁶**

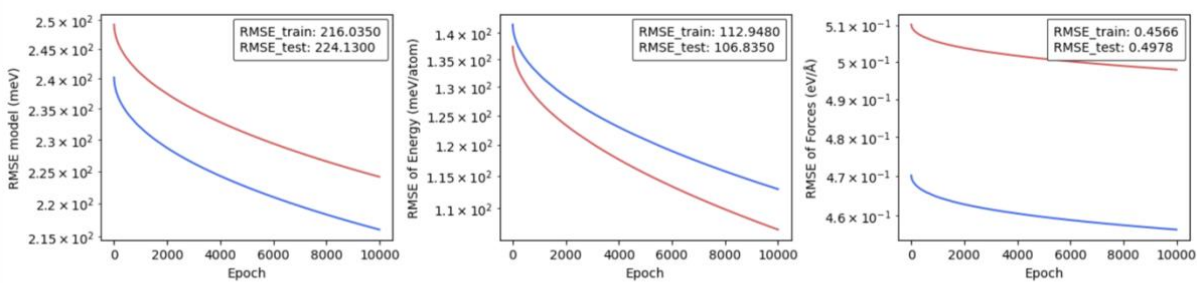

**LR = 10⁻⁵**

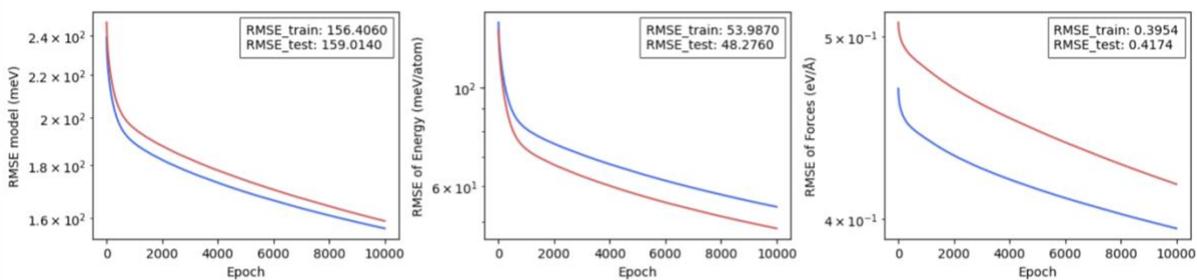

**LR = 10⁻⁴**

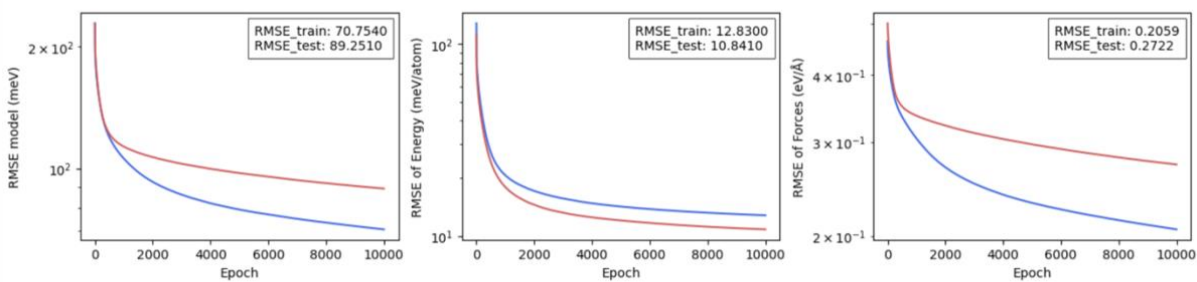

**LR = 10⁻³**

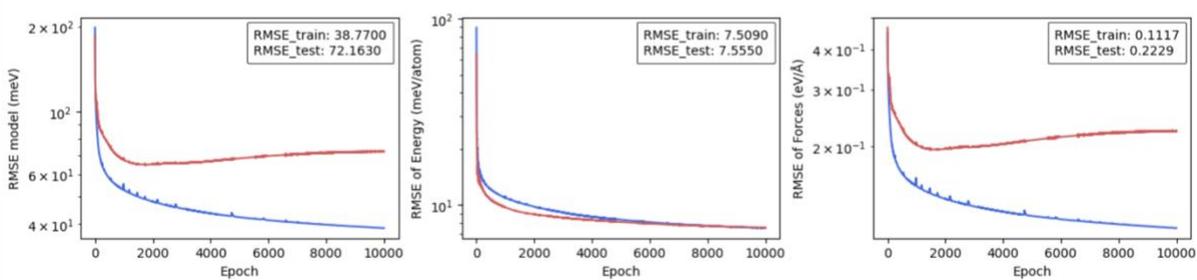

**LR = 10⁻²**

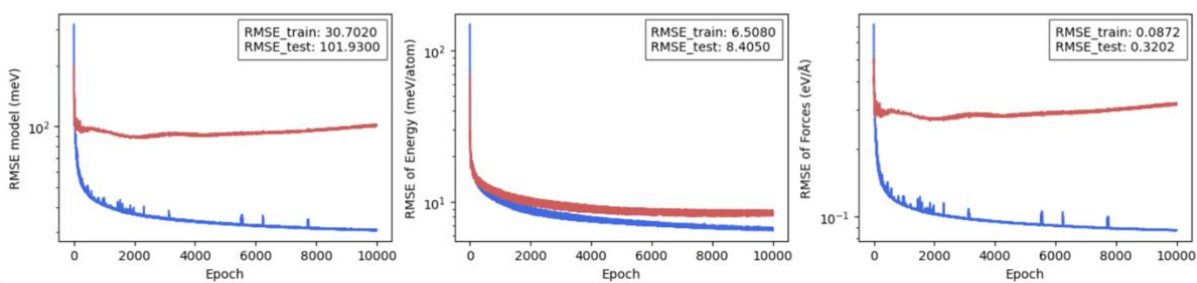

**Figure S8.** The overall prediction convergence (including energy and force) during the MLP training process for the OC20 subset using the Adagrad optimizer with LR from $10^{-2}$ to $10^{-6}$.

**LR = 10⁻⁶**

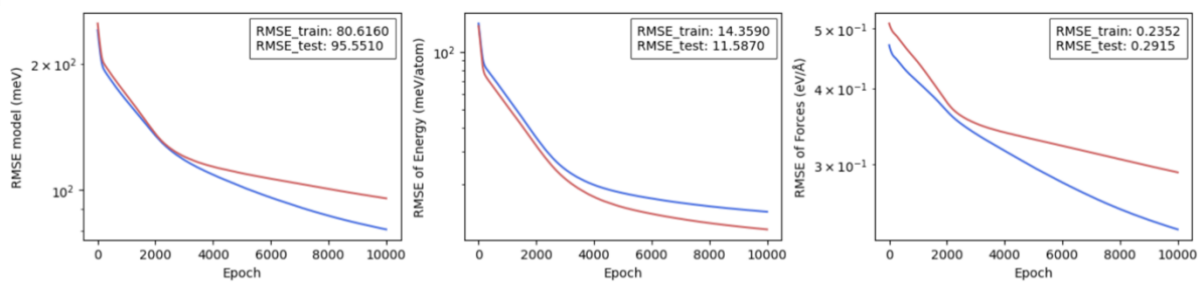

**LR = 10⁻⁵**

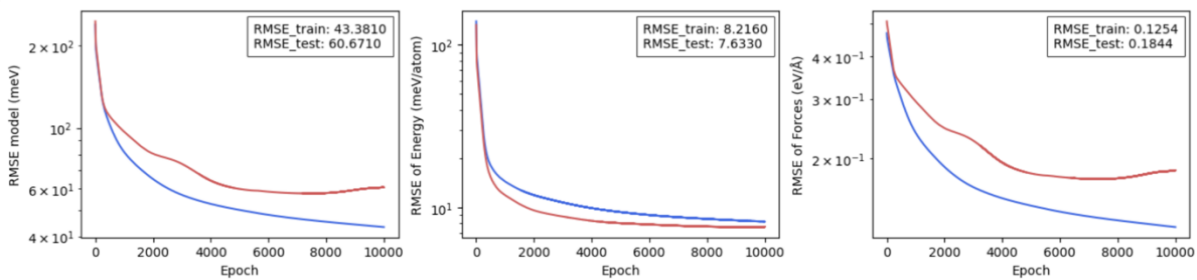

**LR = 10⁻⁴**

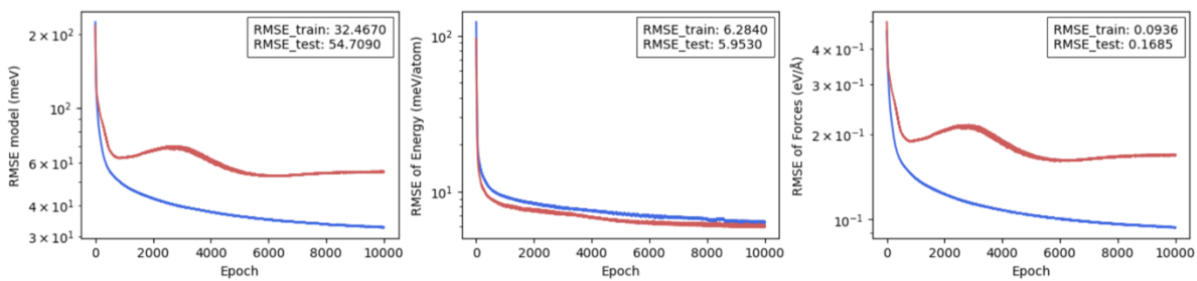

**LR = 10⁻³**

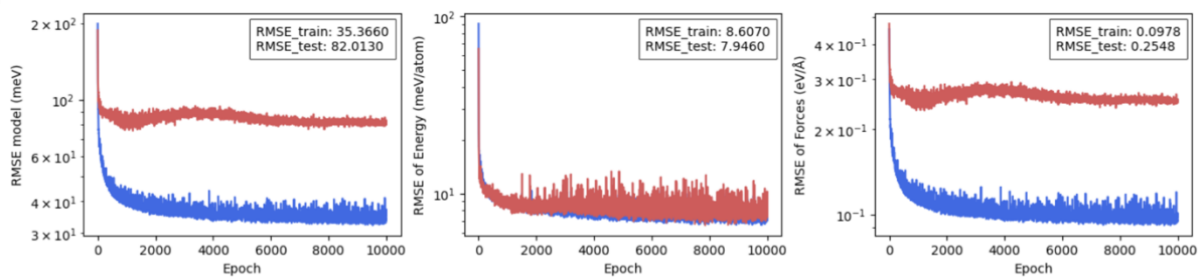

**LR = 10⁻²**

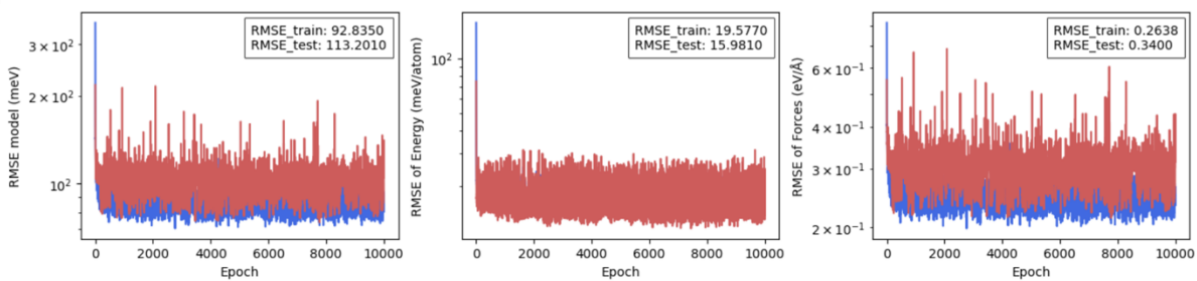

**Figure S9.** The overall prediction convergence (including energy and force) during the MLP training process for the OC20 subset using the Adamax optimizer with LR from $10^{-2}$ to $10^{-6}$.

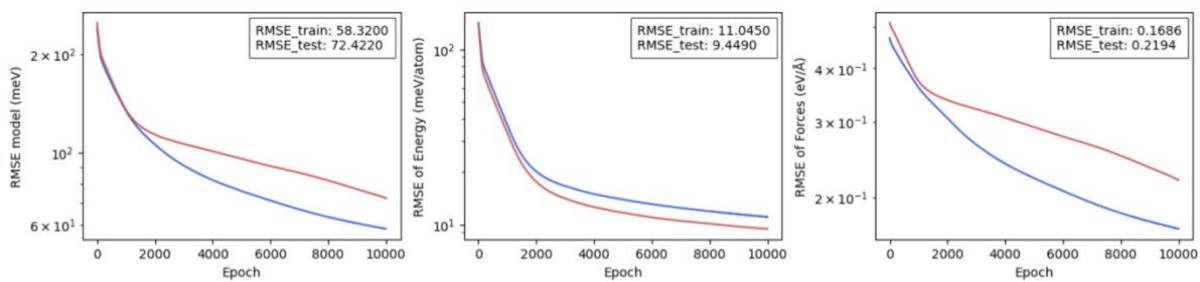
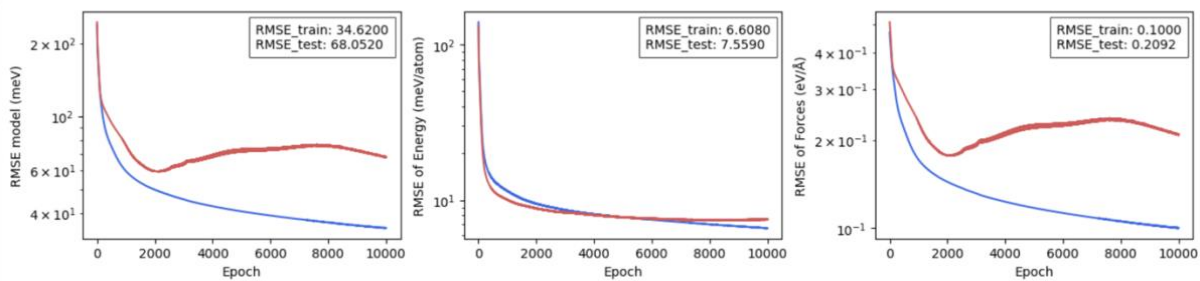
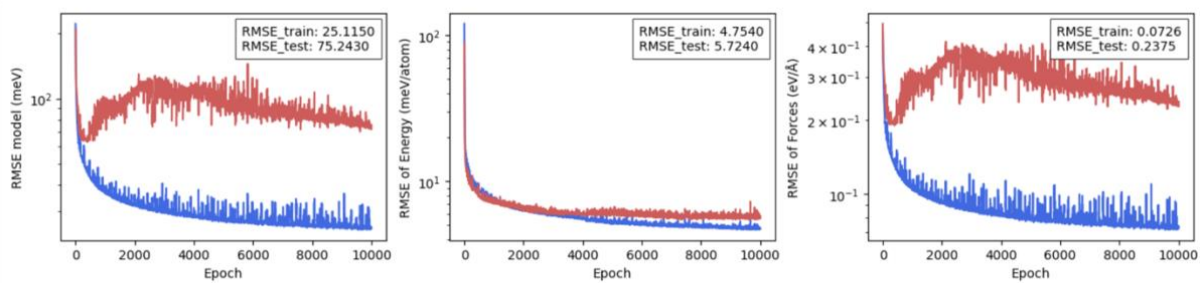
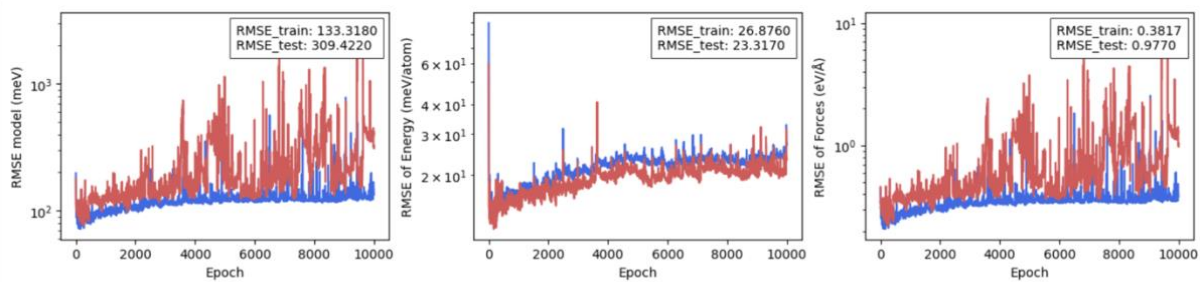
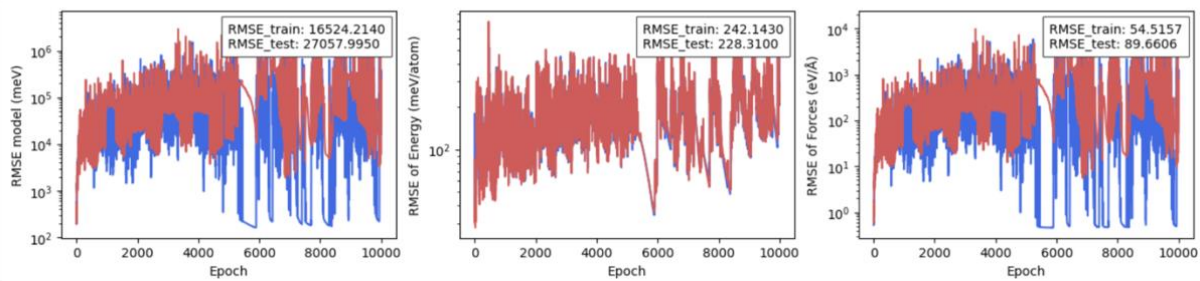

**Figure S10.** The overall prediction convergence (including energy and force) during the MLP training process for the OC20 subset using the Adamw optimizer with LR from $10^{-2}$ to $10^{-6}$.